\documentclass[a4paper,twocolumn,11pt,accepted=2023-02-25]{quantumarticle}
\pdfoutput=1
\usepackage[utf8]{inputenc}
\usepackage[english]{babel}
\usepackage[T1]{fontenc}

\usepackage{cite}

\usepackage{tikz}
\usepackage{lipsum}

\usepackage{graphicx,subfigure}
\usepackage{comment}
\usepackage{float}
\usepackage{color}
\usepackage{epstopdf}
\usepackage{amsmath}
\usepackage{amsthm}
\usepackage{amssymb}
\usepackage{amsfonts}
\usepackage{txfonts}
\usepackage{ulem}
\usepackage{mathtools}
\usepackage{bm}

\usepackage{graphicx}
\usepackage{ulem}
\usepackage{dcolumn}
\usepackage{epsfig}
\usepackage{bm}
\usepackage{array}
\usepackage{hyperref}

\newcommand{\new}[1]{\textcolor{black}{#1}}

\newcommand{\NT}{N}
\newcommand{\bra}[1]{\ensuremath{\langle#1|}}
\newcommand{\ket}[1]{\ensuremath{|#1\rangle}}

\newcommand{\mean}[1]{\ensuremath{\big\langle #1 \big\rangle}}

\newcommand{\makeref}[1]{(\ref{#1})}
\newcommand{\Jx}[1]{\hat{J}_{x}^{\{#1\}}}
\newcommand{\Jy}[1]{\hat{J}_{y}^{\{#1\}}}
\newcommand{\Jz}[1]{\hat{J}_{z}^{\{#1\}}}
\newcommand{\Jp}[1]{\hat{J}_{\perp}^{\{#1\}}}

\newcommand{\vect}[1]{\bm{#1}}

\newcommand{\m}{\nu_E}
\newcommand{\phiampn}{\Lambda_{pn}}

\newcommand{\be}{\begin{equation}}
\newcommand{\ee}{\end{equation}}
\newcommand{\ud}{\mathrm{d}}
\newcommand{\beq}{\begin{eqnarray}}
\newcommand{\eeq}{\end{eqnarray}}

\allowdisplaybreaks

\begin{document}

\title{Quantum-enhanced differential atom interferometers and clocks with spin-squeezing swapping}

\author{Robin Corgier}
\affiliation{QSTAR, INO-CNR and LENS, Largo Enrico Fermi 2, 50125 Firenze, Italy.}
\affiliation{LNE-SYRTE, Observatoire de Paris, Université PSL, CNRS, Sorbonne Université 61 avenue de l’Observatoire, 75014 Paris, France}
\orcid{0000-0002-6846-9249}

\author{Marco Malitesta}
\affiliation{QSTAR, INO-CNR and LENS, Largo Enrico Fermi 2, 50125 Firenze, Italy.}

\author{Augusto Smerzi}
\affiliation{QSTAR, INO-CNR and LENS, Largo Enrico Fermi 2, 50125 Firenze, Italy.}
\orcid{0000-0002-4967-6939}

\author{Luca Pezzè}
\affiliation{QSTAR, INO-CNR and LENS, Largo Enrico Fermi 2, 50125 Firenze, Italy.}
\orcid{0000-0003-0325-9555}

\begin{abstract}
Thanks to common-mode noise rejection, differential configurations are crucial for realistic applications of phase and frequency estimation with atom interferometers. 
Current differential protocols with uncorrelated particles and mode-separable settings reach a sensitivity bounded by the standard quantum limit (SQL).
Here we show that differential interferometry can be understood as a distributed multiparameter estimation problem and can benefit from both mode and particle entanglement.
Our protocol uses a single spin-squeezed state that is mode-swapped among common interferometric modes. 
The mode swapping is optimized to estimate the differential phase shift with sub-SQL sensitivity.
Numerical calculations are supported by analytical approximations that guide the optimization of the protocol. 
The scheme is also tested with simulation of noise in atomic clocks and interferometers.  
\end{abstract}

\maketitle

Atom interferometers (AIs) are extraordinarily sensitive devices~\cite{Berman1997,Cronin09,Tino14} that find key applications in fundamental and applied science~\cite{Safronova18,Bongs19}.
Thanks to the coupling of atoms with external forces and light fields, AIs allows for ultraprecise measurements in inertial sensors~\cite{Geiger20} and atomic clocks~\cite{PoliNC2012, LudlowRMP2015}. 

Absolute measurements with AIs are limited by the onset of systematic effects and/or phase noise, inherent to the interferometer itself and to the environment. 
Therefore, experiments aiming at precision measurements often benefit of a differential configuration.
In this case, two interferometers (A and B), working simultaneously, acquire a total phase shift $\theta_{A,B} = \phi_{A,B} + \phi_{\rm pn}$, respectively, where $\phi_{A,B}$ are the signals and $\phi_{\rm pn}$ encompasses systematic effects or/and stochastic phase noise that are common to both sensors.
Correlations between the output signals of the two interferometers can be used to infer $\phi_A-\phi_B$~\cite{LandiniNJP2014,FosterOL2002,EckertPRA2006,StocktonPRA2007,VaroquauxNJP2009, PereiraPRA2015}, while cancelling the common-mode noise.
Differential AIs configurations have been exploited for gravity gradiometry~\cite{Trimeche19, McGuirk02, Perrin19, Caldani19, Sorrentino14}, 
gravity-field curvature measurements~\cite{RosiPRL2015},
relativistic geodesy~\cite{Philipp20},
the measurements of the gravitational constant~\cite{Rosi14}, to probe the law of gravitation through tests of the universality of free-fall~\cite{SchlippertPRL14,Barrett16,Rosi17,Asenbaum20,Barrett22}, and are also expected to find applications in the detection of gravitational waves \cite{TinoCQG2007, DimopoulosPRD2008, GrahamPRL2013,CanuelCQG20}.
Furthermore, correlation spectroscopy exploits differential frequency measurements.
This technique is based on the simultaneous interrogation of two atoms \cite{ChouPRL2011, ClementsPRL2020} or two atomic ensembles with the same laser: it allows for the cancellation of laser fluctuations and is immune to dead times. 
Differential frequency comparison in atomic clocks has been exploited for the measurement of the gravitational redshift \cite{ChouSCNIENCE2010,BothwellNATURE2022, ZhengNATURE2022} by taking advantage of long interrogation times.  

Current differential AIs use independent interferometers and uncorrelated atoms: a distributed sensing configuration commonly indicated as mode-separable and particle-separable \new{(MsPs)}~\cite{GessnerPRL2018}.
In this case, the sensitivity of the differential measurement is inherently limited by the standard quantum limit (SQL)~\cite{Gauguet09, Sorrentino14, Rosi14,GessnerPRL2018, LiuNATPHOT2021,JanvierPRA22}.
This bound is given by $\Delta^2 (\theta_A - \theta_B)_{\rm SQL} = 4/\NT$, where $\NT$ is the total number of particles~\cite{GessnerPRL2018, notaSQL}. 
A promising research goal is to devise feasible schemes that reduce the quantum noise in interferometric measurements and, in particular, overcome the SQL with the use of quantum resources \cite{Pezze18, Szigeti21}.

The experimental investigation on entanglement-enhanced AIs has mainly focused on the generation and the manipulation of spin-squeezed states~\cite{Pezze18}.
Spin squeezing in momentum modes, suitable for inertial sensing with AIs, can be created through atom-atom interaction in Bose-Einstein condensates~\cite{PRLSzigeti20, PRACorgier21, PRLCorgier21} or atom-light interaction in an optical cavity~\cite{PRLSalvi2018, GreveARXIV}.
Recent experimental achievements include the creation of useful entangled states through the transfer of squeezing generated in internal states into well-defined and separated external momentum modes~\cite{AndersPRL2021}, the
demonstration of long-lived spin-squeezed states~\cite{ArxivHuang20}, and the implementation of a full spatial interferometer scheme with a direct observation of sub-SQL sensitivity~\cite{GreveARXIV}.
Furthermore, entanglement-enhanced atomic clocks using spin-squeezed states have been demonstrated~\cite{LouchetChauvetNJP2010, KrusePRL2016, Pedrozo20, MaliaPRL2020}.

\begin{figure*}[ht!]\includegraphics[width=\textwidth]{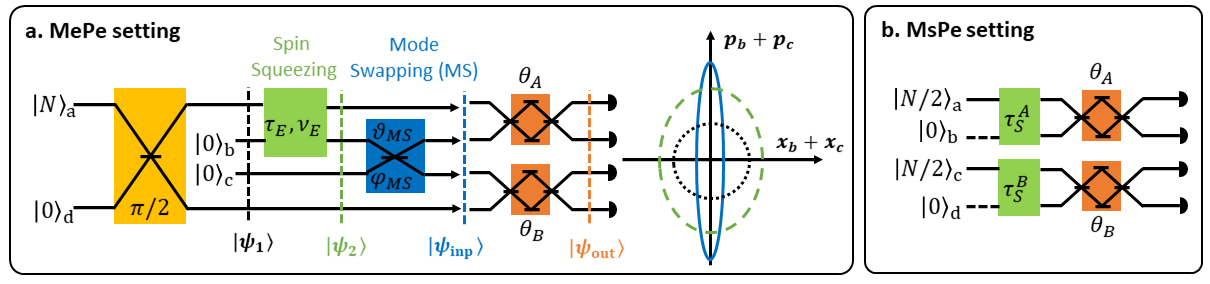}
\caption{Differential interferometry within (a) mode-entangled particle-entangled (MePe) and a (b) mode-separable particle-entangled (MsPe) settings. 
Panel (a): A single input atomic ensemble of $N$ atoms is initially coherently split between modes $a$ and $d$ by a first $\pi/2$ pulse (yellow box). 
A spin-squeezed state is generated in modes $a$ and $b$ via one-axis twisting evolution of strength $\tau_E$ \new{and orientation angle $\nu_E$} (green box, \new{see Eq.~\ref{Eq_State_step_2_OAT}}). 
The MS (blue box) distributes the spin squeezing between the modes $b$ and $c$ \new{through the parameters $\vartheta_{\rm MS}$ and $\varphi_{\rm MS}$ representing the strength and the phase of the coupling laser (see Eq.~\ref{psinp})}. 
Two separated Mach-Zehnder interferometers (orange boxes) are used to estimate $\theta_A$ and $\theta_B$.
The inset shows a schematic of the quantum state at different stages of the protocol on the $(x_b+x_c)$-$(p_b+p_c)$ quadrature plane, obtained within a Holstein-Primakoff transformation (see text).
The distribution corresponding to the state $\ket{\psi_2}$ has variances $(1+e^{-2r})/2$ and $(1+e^{+2r})/2$ along the $x_b+x_c$ and $p_b+p_c$ direction, respectively (green dashed line), where $r=N\tau_E/4$. 
These are wider than the isotropic vacuum variance equal to 1 (dotted black line). 
In contrast, the state after MS, $\ket{\psi_{\rm inp}}$ has variance (blue solid line) $e^{-2r}$ along $x_b+x_c$ and $e^{+2r}$ along $p_b+p_c$, below the vacuum limit. 
Panel (b): Two separated interferometers, using spin squeezed states of $N/2$ atoms ($\tau_S^{A,B}$ indicates the squeezing strength -- green boxes).
In both schemes, the $\theta_A$ and $\theta_B$ are evaluated locally by counting the number of atoms at each output detectors.}
\label{fig_1}
\end{figure*} 

In this manuscript, we propose an efficient method for quantum-enhanced differential phase estimation that exploits  spin-squeezing.
A spin squeezed state is first generated between the two modes of the interferometer A, see Fig.~\ref{fig_1}(a), for instance via one-axis twisting~\cite{KitagawaPRA1993}.
The squeezed state is then shared with the modes of interferometer B via a linear transformation  -- namely an optimized beam-splitter -- and mixed with a coherent spin state.
We indicate this operation as mode-swapping (MS): it generates entanglement between the modes of the two interferometers, thus realizing a mode-entangled and particle-entangled (MePe) distributed quantum sensor~\cite{GessnerPRL2018}.
This scheme extends to atomic systems a protocol proposed in Ref.~\cite{MalitestaARXIV}, which uses optical squeezed-vacuum and coherent states in a network of Mach-Zehnder sensors.
With {\it ab-initio} numerical calculations supported by an analytical approach, we {\it i}) predict the possibility to overcome the SQL with a substantial sensitivity gain, up to $\Delta^2(\theta_A-\theta_B) = O(1/N^{3/2})$, and {\it ii}) demonstrate that the protocol benefits of common-noise suppression, while using standard measurements and estimation.
The sensitivity scaling $O(1/N^{3/2})$ is due to the one-axis-twisting method and can be pushed to the Heisenberg limit by using other spin-squeezing generation schemes. 
It should be emphasized that, with the same resources (namely a coherent and a single spin-squeezed state) in a mode-separable and particle-entangled (MsPe) setting~\cite{GessnerPRL2018}, see Fig.~\ref{fig_1}(b), it is only possible to overcome the SQL by a unit factor.
The practical advantage of our AI protocol is that it fully exploits a single spin-squeezed state for differential measurements: we can thus envisage the realization of high sensitivity enhancement within state-of-the-art capabilities~\cite{Hosten16, CoxPRL2016, Pezze18}, with a variety of applications.

\section{Results}

In the following, we first introduce the interferometer scheme of Fig.~\ref{fig_1}(a), see Sec.~\ref{interf}. 
The sensitivity of the device is then evaluated using an analytical approach, in Sec.~\ref{HP}, as well as numerical calculations for collective spin systems, in Sec.~\ref{Numerical}. 
Finally, in Sec.~\ref{Examples}, we provide explicit examples of differential phase and frequency estimation in the presence of common noise, discussing the advantages with respect to mode-separable approaches.  

\subsection{Interferometer scheme}
\label{interf}

The differential scheme of Fig.~\ref{fig_1}(a) starts with $\NT$ atoms prepared in mode $a$, while the other modes ($b$, $c$ and $d$) are empty at this stage. 
A $50$-$50$ beam splitting pulse between modes $a$ and $d$ [yellow box in Fig.~\ref{fig_1}(a)] generates a binomial coherent particle distribution,
\be \label{psi1}
\ket{\psi_1} = \frac{1}{2^{\frac{N}{2}}} \sum_{m=0}^{\NT} \binom{\NT}{m}^{\frac{1}{2}}~\ket{\NT-m}_a\ket{0}_b\ket{0}_c\ket{m}_d,
\ee
with $\mean{\hat{N}_a} = 
\mean{\hat{N}_d} = \NT/2$ and 
$(\Delta \hat{N}_a)^2 = (\Delta \hat{N}_d)^2 = \NT/2$. 
We then apply a two-mode spin-squeezing transformation to generate entanglement between particles in modes $a$ and $b$ [green box in Fig.~\ref{fig_1}(a)].
We consider one-axis twisting dynamics~\cite{KitagawaPRA1993} followed by an opportune rotation of the state,
\beq
\label{Eq_State_step_2_OAT}
\ket{\psi_2} &=& \exp\left\{-i\m \Jz{a,b}\right\}\nonumber \\
&&\times\exp\left\{-i \tau_E \left(\Jx{a,b}\right)^2\right\} \ket{\psi_1}.
\eeq
\new{In Eq.~\ref{Eq_State_step_2_OAT}, the term $(\Jx{a,b})^2$ is associated to a non-linear evolution due to atom-atom interaction in a Bose-Einstein condensate~\cite{PRLSzigeti20, PRACorgier21, PRLCorgier21} or atom-light interaction in a cavity~\cite{LerouxPRL2010, PRLSalvi2018, GreveARXIV}, and generates a two-mode spin-squeezing dynamics, where $\tau_E$ denotes the effective squeezing time.
The term $\exp\{-i\m \Jz{a,b}\}$ rotates the state in the Bloch sphere by the angle $\m$. 
It is used to optimize the orientation of the quadrature of the quantum state to be sensitive to a phase measurement. 
Experimentally, such transformation can be generated by manipulating the phase of the laser coupling mode $a$ and $b$.}
Here, $\Jx{a,b} = (\hat{a}^\dag \hat{b} + \hat{b}^{\dag}\hat{a})/2$, $\Jy{a,b}= (\hat{a}^\dag \hat{b} - \hat{b}^{\dag}\hat{a})/2i$ and $\Jz{a,b} = (\hat{a}^\dag \hat{a} - \hat{b}^{\dag}\hat{b})/2$ 
($\Jx{c,d} = (\hat{c}^\dag \hat{d} + \hat{d}^{\dag}\hat{c})/2$, $\Jy{c,d} = (\hat{c}^\dag \hat{d} - \hat{d}^{\dag}\hat{c})/2i$ and $\Jz{c,d}= (\hat{c}^\dag \hat{c} - \hat{d}^{\dag}\hat{d})/2$ are collective pseudo-spin operators for the modes $a$ and $b$ ($c$ and $d$). 
The operator $\hat{a}$ ($\hat{b}$, $\hat{c}$ and $\hat{d}$) annihilates a particles in mode $a$ ($b$, $c$, and $d$), while $\hat{a}^\dag$ ($\hat{b}^\dag$, $\hat{c}^\dag$ and $\hat{d}^\dag$) is the corresponding bosonic creation operator.

The key ingredient of our proposal is the MS [blue box in Fig.~\ref{fig_1}(a)], corresponding to a linear coupling between modes $b$ and $c$. 
We indicate as $\ket{\psi_{\rm inp}}=\hat{U}_{\rm MS} \ket{\psi_2}$ the quantum state at this stage, with
\beq 
\label{psinp}
\hat{U}_{\rm MS}&=& \exp\Big\{-i \vartheta_{\rm MS}\cos(\varphi_{\rm MS})\Jx{b,c} \nonumber \\
&&\quad\quad\,\,\,\,\,+i\vartheta_{\rm MS}\sin(\varphi_{\rm MS})\Jy{b,c}\Big\},
\eeq
where $\vartheta_{\rm MS}$ and $\varphi_{\rm MS}$ are, respectively, the strength and the phase of the laser coupling mode $b$ an $c$.
As schematically shown in the inset of Fig.~\ref{fig_1}(a) and explained in more details below, the MS swaps the spin squeezing from the modes $a$ and $b$ (where the squeezing is generated) to a combination of suitable interferometer's modes. 
Without the MS, e.g. \new{coupling strength} $\vartheta_{\rm MS}=0$, the spin squeezing generated in the $a$ and $b$ modes is only useful to estimate $\theta_A$. 
\new{In this study, the combination of a non-linear dynamic of strength $\tau_E$ with an adequate orientation of the four-mode state provided by $\nu_E$ and $\varphi_{MS}$ enables the creation of a quantum-enhanced sensitive states for differential measurements.}

The preparation of $\ket{\psi_{\rm inp}}$ is followed by two independent Mach-Zehnder interferometers, indicated as $A$ (for modes $a$ and $b$) and $B$ (for modes $c$ and $d$). 
The scheme ends with the measurement of the number of particles in each output port, the output state being  $\ket{\psi_{\rm out}}  = \exp(-i \theta_A \Jy{a,b})\, \otimes\, \exp(-i \theta_B \Jy{c,d})\, \ket{\psi_{\rm inp}}$.
The difference $\phi_A-\phi_B$ is estimated as $\theta_A^{({\rm est})} -\theta_B^{({\rm est})}$, where $\theta_A^{({\rm est})}$ and $\theta_B^{({\rm est})}$ are obtained via the standard method of moments applied to each interferometer, separately. 
The corresponding uncertainty, is~\cite{GessnerNATCOMM2020}
\beq \label{DthetaAB}
\Delta^2(\theta_A-\theta_B) &=& \Delta^2\theta_A+\Delta^2\theta_B \\
&-&\dfrac{2\,\Gamma_{AB}}{\left(\dfrac{\partial \langle \Jz{a,b} \rangle}{\partial \theta_A}\right)\left( \dfrac{\partial \langle \Jz{c,d} \rangle}{\partial \theta_B}\right)},\nonumber
\eeq
where expectation values are calculated on the output state $\ket{\psi_{\rm out}}$, $\Delta^2\theta_A = \Delta^2 \Jz{a,b}/(d \mean{\Jz{a,b}}/d \theta_A)^2$, $\Delta^2\theta_B = \Delta^2 \Jz{c,d}/(d \mean{\Jz{c,d}}/d \theta_B)^2$, and the covariance $\Gamma_{AB}=\langle \Jz{a,b} \Jz{c,d} \rangle-\langle \Jz{a,b} \rangle\langle \Jz{c,d} \rangle$ quantifies correlations between measurement observables.
These correlations are engineered by the MS transformation.
At mid-fringe position $\theta_A=\theta_B=\pi/2$ (which, as discussed below is the optimal working point), Eq. (\ref{DthetaAB}) can be rewritten as 
\beq
\label{DthetaABpi2}
&&\Delta^2(\theta_A-\theta_B)\big\vert_{\theta_A=\theta_B=\frac{\pi}{2}} \nonumber \\
&&\quad = \Delta^2\left[ \frac{\Jx{a,b}}{\mean{ \Jz{a,b}}_{\rm inp}} - \frac{\Jx{c,d}}{\mean{\Jz{c,d}}_{\rm inp}}\right]_{\rm inp}.
\eeq
For two uncoupled interferometers, we have $\Gamma_{AB} = 0$ and Eq.~(\ref{DthetaAB}) recovers the sum of uncertainties $\Delta^2(\theta_A-\theta_B) = \Delta^2\theta_A+\Delta^2\theta_B$.
Instead, if the term proportional to $\Gamma_{AB}$ \new{is} positive, then $\Delta^2(\theta_A-\theta_B) < \Delta^2\theta_A+\Delta^2\theta_B$.
In particular, this opens the possibility to overcome the SQL for the estimation of the differential phase.
Through this paper,
\be \label{SQLintro}
\mathcal{G}^2=\dfrac{\Delta^2(\theta_A-\theta_B)_{\rm SQL}}{\Delta^2(\theta_A-\theta_B)}.
\ee
quantifies the sensitivity gain over the SQL~\cite{notaSQL}.

\subsection{Analytical model}
\label{HP}

The analytical description of the interferometer is based on two approximations, see App.~\ref{AppA1} for details.
First, we write the state $\ket{\psi_1}$, Eq. (\ref{psi1}), as $\ket{\psi_1} \approx \ket{{\rm C}(\alpha)}_a \ket{0}_b \ket{0}_c \ket{{\rm C}(\alpha)}_d$,
where $\ket{{\rm C}(\alpha)}$ is a single-mode coherent state of $\alpha^2 = \NT/2$ particles in average \cite{BarnettBOOK}. 
\new{This approximation assumes $\NT \gg 1$, see App.~\ref{AppA1}.
Experimentally, one can easily generate Bose-Einstein condensate with $N$ approximately ranging from $10^2$ to $10^6$ atoms such that the condition $N\gg1$ is easily satisfied.
The phases of the two coherent states in modes $a$ and $d$
are equal and set to zero without loss of generality.}
Furthermore, we describe the one-axis twisting transformation by using a Holstein–Primakoff transformation replacing the operators $\hat{a}$ with the classical number $\hat{a} \sim 
\sqrt{\NT/2}$. 
We obtain 
\be\label{squeezing_step}
\ket{\psi_2}
\approx 
\ket{{\rm C}(\alpha e^{-i \m/2})}_a \ket{{\rm S}(\xi)}_b \ket{0}_c \ket{{\rm C}(\alpha)}_d,
\ee
where $\ket{{\rm S}(\xi)}_b = \hat{S}_b(\xi) \ket{0}_b$ is a single-mode squeezed-vacuum state and $\hat{S}_b(\xi) = e^{(\xi^*\hat{b}\hat{b} - \xi \hat{b}^\dag \hat{b}^\dag)/2}$ is the squeezing operator~\cite{BarnettBOOK} with squeezing parameter $\xi =r e^{i\varphi}$.
Here, $r=\NT \tau_E/4$, $\varphi=\pi/2 + \m$ and $\bar{n}_s = \sinh^2 r$ is the average number of particles in the state. 
Equation~(\ref{squeezing_step}) is expected to be accurate if the squeezing dynamics is applied for a sufficiently short time such that the mode $a$ remains essentially undepleted, namely $\NT/2 \gg \bar{n}_s$ \new{(corresponding to $\tau_E\ll\sqrt{8/N}$)}.
Taking the quadrature operator $\hat{x}_b(\lambda) = (\hat{b}e^{-i\lambda} + \hat{b}^\dag e^{i \lambda})/\sqrt{2}$,
the state $\ket{{\rm S}(\xi)}_b$ is squeezed along the direction $\lambda=\pi/4+\m/2$ in the quadrature plane, namely $\Delta^2 \hat{x}_b(\pi/4+\m/2) = e^{-2r}/2$, and 
anti-squeezed along the orthogonal direction, $\Delta^2 \hat{x}_b(3\pi/4+\m/2) = e^{2r}/2$.
Using $\Delta^2 \Jp{a,b} \approx \NT \Delta^2 \hat{x}_b(\lambda)/4$ and $\mean{\Jz{a,b}} \approx (\NT/2-\bar{n}_s)/2 \approx \NT/4$, we predict a spin-squeezing parameter
\be \label{Eq.squeezing}
\xi_R^2 = \frac{(\NT/2) \min_{\perp} \Delta^2 \Jp{a,b}}{\mean{\Jz{a,b}}^2} = 2 \min_{\lambda} \Delta^2 \hat{x}_b = e^{-2r}.
\ee
This expression agrees with the exact result~\cite{KitagawaPRA1993}, up to the third order in $\NT\tau_E$~\cite{SorelliPRA2019}, namely $e^{-2r} = 1- \NT\tau_E/2 + (\NT\tau_E)^2/8
+ O(\NT\tau_E)^3$. 
Finally, the MS operation consists of a linear coupling between modes $b$ and $c$.
It is described by the unitary matrix
\begin{align}\label{U_matrix}
    \begin{pmatrix}\hat{U}^{\dagger}_{\rm MS}\hat{b}\hat{U}_{\rm MS}\\\hat{U}^{\dagger}_{\rm MS}\hat{c}\hat{U}_{\rm MS}\end{pmatrix} = \begin{pmatrix}|u_{bb}|&-|u_{cb}|e^{-i\delta_{cb}}\\|u_{cb}|e^{i\delta_{cb}}&|u_{bb}|\end{pmatrix} 
    \begin{pmatrix}\hat{b}\\\hat{c}\end{pmatrix},
\end{align}
with $\hat{U}_{\rm MS}$ given by  Eq. \makeref{psinp}~\cite{note2}.
The state at the input ports of the two interferometers writes
\be\label{psinpHP}
\ket{\psi_{\rm inp}}
\approx 
\ket{{\rm C}(\alpha e^{-i \m/2})}_a \left( \hat{U}_{\rm MS} \ket{{\rm S}(\xi)}_b \ket{0}_c \right)\ket{{\rm C}(\alpha)}_d.
\ee
where the unitary transformation\new{, $\hat{U}_{\rm MS}$,} applies to the state $\ket{{\rm S}(\xi)}_b \ket{0}_c$.

Based on the approximations discussed above, we have thus established a relation, in the appropriate regime, between an atomic system using spin-squeezed states and a quantum optical system using squeezed vacuum states, the MS playing the role of a quantum circuit.
We can thus apply the results of Ref.~\cite{MalitestaARXIV}, which discussed the sensitivity of a distributed quantum sensor of Mach-Zehnder interferometers using coherent and single-mode squeezed states.   
Taking Eq.~(\ref{psinpHP}) as sensor's input state, we can obtain an analytical expression for Eq.~(\ref{DthetaAB})~\cite{MalitestaARXIV}, which is lengthy and explicitly given in App.~\ref{AppA2}.
In particular, it reads $\Delta^2(\theta_A-\theta_B)=\Delta^2(\theta_A-\theta_B)\big\vert_{\theta_A = \theta_B = \pi/2}+\mathcal{Q}(\cot{\theta_A},\cot{\theta_B})$,
where $\mathcal{Q}\geq 0$ is a second-degree polynomial \new{function} in the variables $\cot{\theta_A}$ and $\cot{\theta_B}$.
In the rest of this section, we restrict to the optimal working point $\theta_A=\theta_B=\pi/2$, where $\mathcal{Q}=0$.

To clarify the role played by the MS for highly sensitive differential estimation, we rely again on a Holstein–Primakoff transformation. 
This time, it is applied to the operators $\hat{a}$ and $\hat{d}$ relative to the output modes, \new{that are replaced by the} classical numbers $\hat{a} \sim e^{-i\m/2}\sqrt{\NT/2}$ and $\hat{d} \sim \sqrt{\NT/2}$, respectively. 
As detailed in App.~\ref{AppA3}, Eq.~(\ref{DthetaABpi2}) rewrites in terms of the quadrature variance
\be \label{HPapprox}
\Delta^2(\theta_A-\theta_B)\big\vert_{\theta_A = \theta_B = \pi/2} \approx 4 \frac{\Delta^2(\hat x_b + \hat x_c)}{\NT},
\ee
where $\hat x_b = (\hat{b}e^{i\m/2} + \hat{b}^\dag e^{-i\m/2})/\sqrt{2}$, $\hat p_b = (\hat{b}e^{i\m/2} - \hat{b}^\dag e^{-i\m/2})/\sqrt{2}i$ \new{and $\hat{x}_c=(\hat{c}+\hat{c}^{\dagger})/\sqrt{2}$, $\hat{p}_c=(\hat{c}-\hat{c}^{\dagger})/\sqrt{2}i$.} According to Eq.~(\ref{HPapprox}), the phase uncertainty of the differential measurement is equivalent to the variance of the quadrature operator $\hat x_b+\hat x_c$.
Without mode swapping, namely when considering the state $\ket{{\rm S}(\xi)}_b\ket{0}_c$, we have $\Delta^2(\hat x_b +\hat x_c) = \Delta^2 \hat x_b + \Delta^2 \hat x_c$. 
Hence, whatever amount of squeezing we generate in the $x_b$ quadrature, it will provide limited benefit to the sensitivity of the differential estimation, which is indeed dominated by the vacuum fluctuations in mode $c$: $\Delta^2(\hat x_b +\hat x_c) = (1+e^{-2r})/2$. 
In contrast, it is possible  to reduce the variance of $\hat x_b+\hat x_c$ below the vacuum fluctuations by exploiting the MS transformation Eq.~(\ref{U_matrix}). 
The smallest quadrature variance -- and, thus, the best sensitivity -- is clearly reached when both the rotation angle $\m$ in Eq. \makeref{Eq_State_step_2_OAT} and the parameters of Eq. \makeref{U_matrix} are optimized. We find $\m=-\pi/4+l\pi+4m\pi$ and $\delta_{cb}=-\pi/8+l\pi/2$, with $l,m\in \mathbb{Z}$, and $|u_{bb}|=|u_{cb}|=1/\sqrt{2}$ [see App. \ref{AppB1}]. In the following, we choose $l=-1, m=1$, getting $\m=11\pi/4$ and $\delta_{cb}=-5\pi/8$. Since $\delta_{cb}=\varphi_{MS}-\pi/2$, this corresponds to an optimal MS operation in Eq.~(\ref{psinp}) with $\varphi_{\rm MS} = - \pi/8$, and the condition $|u_{bb}|=|u_{cb}|=1/\sqrt{2}$ translates to $\vartheta_{\rm MS}=\pi/2$.
It is shown in App.~\ref{AppA3} that such optimal choice of parameters yields $\Delta^2 (\hat x_b+\hat x_c)=e^{-2r}$: 
the squeezing initially generated in the quadrature plane $\hat x_b$-$\hat p_b$ has been ``swapped" to the $\hat x_b+ \hat x_c$ quadrature  [see the inset of Fig.~\ref{fig_1}(a)] by means of a suitable MS operation.

According to Eq.~\makeref{HPapprox}, $\Delta^2(\hat x_b+\hat x_c)=e^{-2r}$ corresponds to an estimation uncertainty
$\Delta^2(\theta_A-\theta_B) = 4e^{-2r}/\NT$, which predicts a sensitivity gain
\be \label{gaine2r}
\mathcal{G}^2=e^{2r},
\ee
below the SQL.
Therefore, through MS operation, we are able to convert a squeezing useful for single-parameter estimation to a squeezing useful to beat the SQL in the differential measurement. 
\new{Without relying to a Holstein-Primakoff transformation, we obtain }
\be \label{nsggN}
\mathcal{G}^2 = (e^{-2r} + \bar{n}_s/\NT)^{-1},
\ee
\new{which is valid for $\NT \gg \bar{n}_s$.
The detailed derivation of Eq.~(\ref{nsggN}) can be found in App.~\ref{AppB1}.}
For sufficiently small values of $\tau_E$, such that $\NT \gg e^{2r}\bar{n}_s \approx 4\bar{n}_s^2$, the  term $\bar{n}_s/\NT$ in Eq.~\makeref{nsggN} can be neglected, recovering the gain factor Eq.~(\ref{gaine2r}).
Instead, at the optimal working point $\bar{n}_s=\sqrt{\NT}/2$, which still satisfies the condition $\bar{n}_s \ll \NT$ \new{for $N\gg 1$,} Eq.~\makeref{nsggN} predicts a maximum gain 
\be \label{gainmax}
\mathcal{G}^2_{\rm max}=\sqrt{\NT},
\ee
corresponding to the minimum uncertainty $\min_{\bar{n}_s}\Delta^2(\theta_A-\theta_B)\big\vert_{\theta_A = \theta_B = \pi/2}=4/\NT^{3/2}$.

\subsection{Numerical results}
\label{Numerical}

In the following, we relieve the approximations used in the previous section and perform numerical computation of collective spin-coupling operations. 
In particular, we provide a detailed analysis of Eqs.~(\ref{DthetaAB})-(\ref{SQLintro}) as well as a comparison with the MsPe scheme of Fig.~\ref{fig_1}(b). 
For the sake of clarity, we distinguish the squeezing strength $\tau_E$ of the MePe strategy of Fig.~\ref{fig_1}(a), to the one of MsPe strategy, $\tau_S^A$ and $\tau_S^B$ of Fig.~\ref{fig_1}(b). 
The squeezing strengths are normalized over the value $\tau_{\rm ref}=1.2 (\NT/2)^{-2/3}$ providing the optimal amount of spin squeezing within the one-axis-twisting scheme~\cite{KitagawaPRA1993} with $N/2$ atoms. 

\subsubsection{Gain and bandwidth}

\begin{figure}[t!]\includegraphics[width=\columnwidth]{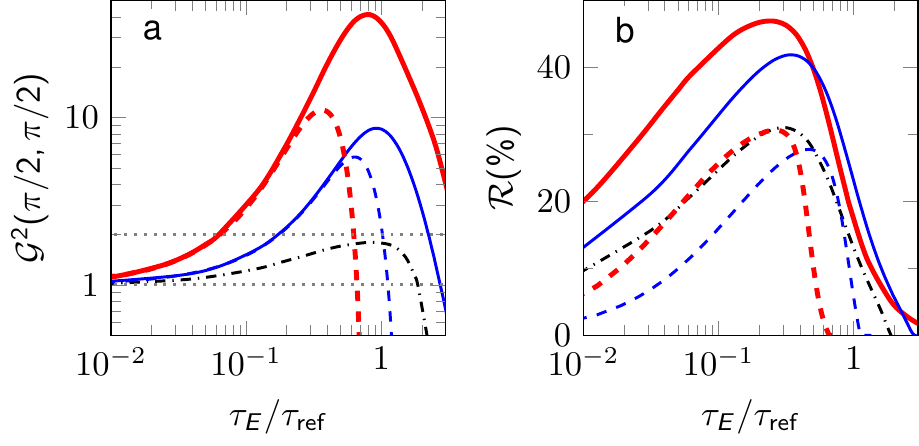}
\caption{(a) Sensitivity gain, $\mathcal{G}^2(\pi/2,\pi/2)$, and (b) bandwidth, $\mathcal{R}$, as a function of the squeezing strength $\tau_E$.
In both panels, the thin blue (red thick) solid line show numerical results obtained for the MePe setting and total $N=100$ ($N=2000$) particles.
The corresponding dashed lines are analytical predictions. 
\new{The black dot-dashed line is obtained for the MsPe setting with $N=100$, using a spin squeezed state ($\tau_S^A=\tau_E$) and a coherent spin state ($\tau_S^B=0$), without MS. 
A similar behavior (not shown) is obtained for $N=2000$.}}
\label{fig_2}
\end{figure}

In Fig~\ref{fig_2}(a), we plot the sensitivity gain, $\mathcal{G}^2$, at the optimal working point $\theta_A=\pi/2$ and $\theta_B=\pi/2$ (corresponding to the maximum slope point of the Ramsey fringes for both interferometers), as a function of the squeezing time $\tau_E/\tau_{\rm ref}$.
The different solid lines are obtained for different number of particles, $N=100$ (thin blue) and $N=2000$ (thick red), with $N/2$ particles entering each interferometer. 
For relatively small values of $\tau_E$, \new{e.g. $\tau_E/\tau_{\rm ref}\lesssim 0.6$ for $N=100$ and $\tau_E/\tau_{\rm ref}\lesssim 0.2$ for $N=2000$ in agreement with $\tau_E\ll\sqrt{8/N}$}, the numerics follow the analytical prediction of Eq.~(\ref{nsggN}), shown as the dashed lines.
The spin-squeezed state is optimally rotated by $\m$ around $\Jz{a,b}$ while, for the MS operation, we have used the optimized laser strength, $\vartheta_{\rm MS} = \pi/2$, corresponding to a 50-50 beam-splitter, with a fixed (not-optimized) laser phase, $\varphi_{\rm MS} = \pi/2$, in Eq.~(\ref{psinp}). 
As expected from Eqs.~(\ref{HPapprox})-(\ref{nsggN}), we observe a substantial amount of gain over the SQL, despite the phase of the MS, $\varphi_{\rm MS}$, is not optimized here (see discussion below). 
In addition, at the optimal working point (as a function of $\tau_E$), we obtain a scaling $\mathcal{G}^2_{\rm max} \sim \sqrt{N}$, in agreement with the analytical predictions of Eq.~(\ref{gainmax}).
The dot-dashed line in the figure shows the MsPe case where the MS is not applied ($\vartheta_{\rm MS} = 0$). 
This situation is equivalent to $\tau_{S}^A=\tau_E$
and $\tau_{S}^B=0$ in Fig.~\ref{fig_1}(b): only the estimation of $\theta_A$ benefits from squeezing ($\Delta^2\theta_A < 2/\NT$), while $\theta_B$ is estimated with uncertainty $\Delta^2\theta_B = 2/\NT$.
In this case, we have only a modest gain in the estimation of $\theta_A-\theta_B$ reaching, at best, approximately a factor 2 over the SQL (horizontal dotted line, $\mathcal{G}^2 = 2$)~\cite{notaSS}.

Besides the sensitivity gain, another relevant figure of merit is the relative width of the phase domain $\theta_A,\theta_B\in[0, \pi]$ such that $\mathcal{G}^2(\theta_A,\theta_B)>1$.
We indicate this quantity as the bandwidth, $\mathcal{R}$, defined as 
\be
\label{Eq_Robustness}
\mathcal{R} = \frac{1}{\pi^2}   \int_{\mathcal{G}^2(\theta_A,\theta_B)> 1}
\ud \theta_A \ud \theta_B.
\ee
A plot of $\mathcal{R}$, as a function of $\tau_E/\tau_{\rm ref}$ and different values of $\NT$, is shown in Fig.~\ref{fig_2}(b). 
The different lines correspond to the same parameters as the ones in Fig.~\ref{fig_2}(a).
Although the analytical predictions (shown as by dashed lines) preserve the qualitative trend, in this case, they do not reproduce the numerical findings.
\new{The difference here is due to the slight difference of the statistic in the initial states due to the approximations made: details can be found in Appendix~\ref{AppA1} and \ref{AppC}.}
Overall, we see that increasing the number of particles from $N=100$ to $2000$ leads to a larger optimal sensitivity gain, and more surprisingly, to an improved robustness.
The MS clearly improves the performance of the differential interferometer, when compared to the mode-separable case, using a single spin-squeezed state.

\begin{figure}[t!]\includegraphics[width=\columnwidth]{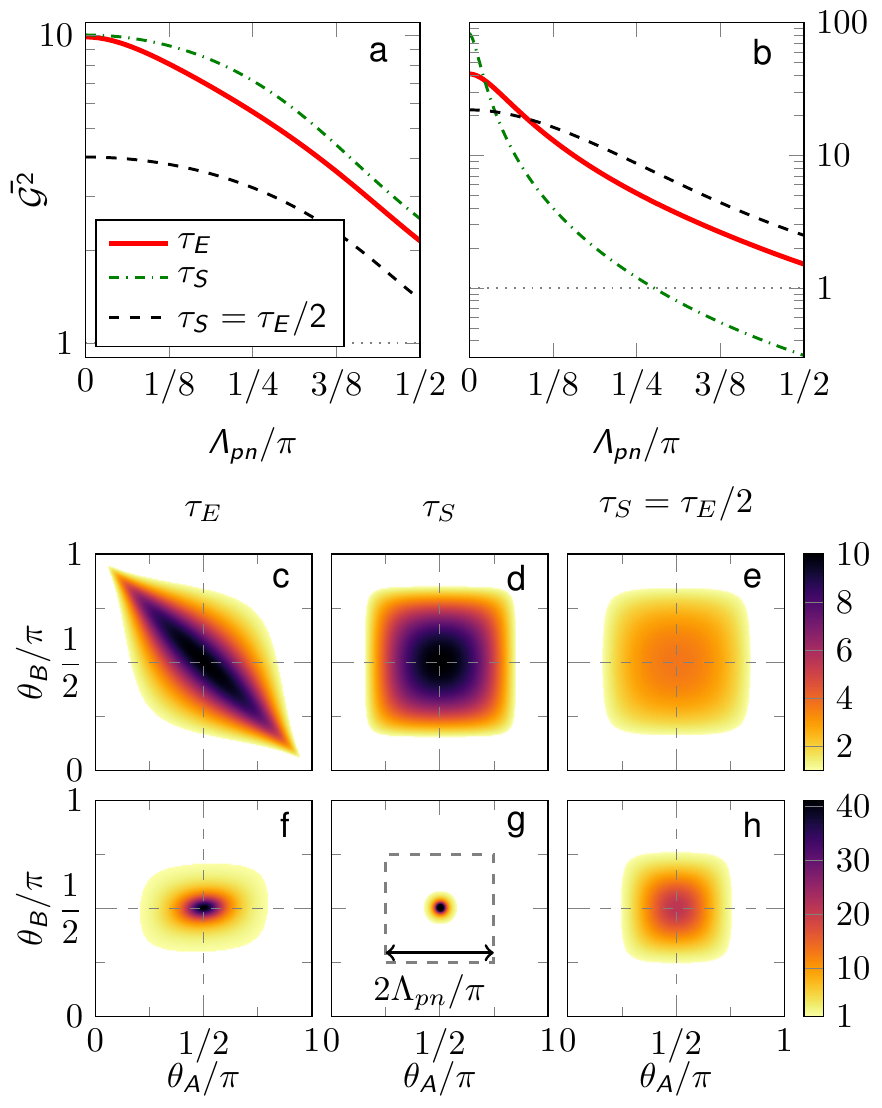}
\caption{Average sensitivity gain, $\bar{\mathcal{G}^2}$, Eq. (\ref{Gave}), as a function of the noise amplitude $\phiampn$ for different squeezing strength corresponding to 10\,dB gain (a) and maximum gain (b). 
The solid red line is the MePe with squeezing strength $\tau_E/\tau_{\rm ref}=0.25$ (a) and $\tau_E/\tau_{\rm ref}=0.75$ (b). 
The green dot-dashed line is the corresponding MsPe strategy using two squeezed states with $\tau_S^A=\tau_S^B =\tau_S= 0.24~\tau_{\rm ref}$ (a) and $\tau_S/\tau_{\rm ref}=1$ (b). 
For comparison, the black dashed line is the MsPe strategy with $\tau_S= \tau_E/2$.  
Panel (c-h): Sensitivity gain, $\mathcal{G}^2(\theta_A,\theta_B)$ (color plot), as a function of $\theta_A$ and $\theta_B$. 
Panel (c-e) (resp. (f-h)) correspond to the case shown in panel (a) (resp. (b)). From left to right: mode-entanglement strategy with squeezing strength $\tau_E$, (c) and (f); mode-separable strategy with squeezing strength $\tau_S$, (d) and (g); and  $\tau_E/2$ (e) and (h).
In all panels $N=2000$.}
\label{fig_3}
\end{figure}

Comparing the two panels in Fig~\ref{fig_2}, we observe that the optimal gain point does not corresponds to the point of maximal bandwidth.
To quantify the trade-off between gain and sensitivity bandwidth, we define the average gain in a squared  box of total length $2\phiampn$ centered a $\phi_A=\phi_B=\pi/2$,
\be \label{Gave}
\bar{\mathcal{G}}^2(\phiampn)= \dfrac{\int_{-\phiampn}^{\phiampn}\, d\phi_{\rm pn}\,\mathcal{G}^2(\pi/2+\phi_{\rm pn},\pi/2+\phi_{\rm pn})}{(2\phiampn)^2},
\ee
\new{where $\phiampn$ denotes the maximum phase noise amplitude common to both interferometer shifting the detection from the optimal working point, $\theta_{A,B}=\pi/2$, to a non-optimal one.
Such situation can be obtained experimentally due to common vibration of the plate-form for instance or other bias phase-shift changing randomly shot-to-shot the working point of the differential measurement.}

In Fig~\ref{fig_3}(a) and (b), we plot the MePe gain $\bar{\mathcal{G}}^2(\phiampn)$ for $\NT=2000$ and different values of the squeezing strength (thick red line):
in panel (a), $\tau_E$ is chosen such that $\mathcal{G}^2(\pi/2,\pi/2) = 10$~dB; 
in panel (b), we considered the value of $\tau_E$ that maximizes $\mathcal{G}^2(\pi/2,\pi/2)$.
\new{These two configurations are respectively obtained for $\tau_E/\tau_{\rm ref}=0.25$ (a) and $\tau_E/\tau_{\rm ref}=0.75$ (b).}
The average gain decreases with $\phiampn$ but it remains above 1 for all values of $\phiampn$.  
We also compare the MePe strategy with the MsPe one (green dot-dashed line) using two squeezed states $\tau_S^A = \tau_S^B = \tau_S$ with (a) $\tau_S/\tau_{\rm ref}=0.24$ corresponding to a maximum gain of 10\,dB, and (b) the optimal $\tau_S/\tau_{\rm ref}=1$ value.
In the case (a), the MsPe shows a slightly higher average gain than the MePe scheme, but the former uses two squeezed states instead of one. 
In (b), the average gain of the MsPe case is higher for small $\phiampn$, but it quickly degrades with $\phiampn$ as a consequence of the small bandwidth of highly spin-squeezed states. 
In contrast, the MePe strategy provides a higher average gain for large values of $\phiampn$.
Finally, taking the total squeezing time as a resource, we compare the MePe strategy with the MsPe one with $\tau_S^A = \tau_S^B = \tau_E/2$.
The corresponding average gain is shown as the black dashed line in Fig~\ref{fig_3}(a) and (b).
In the optimal-squeezing case, Fig~\ref{fig_3}(b), the MsPe slightly outperforms the mode-entangled case for sufficiently large $\phiampn$. 
It should be noticed, nevertheless, that higher dynamical range could be obtained at $\phiampn=\pi/2$ for lower value of $\tau_E$ in the MePe strategy (not shown in the graph).

For further clarity, we plot, in Fig~\ref{fig_3}(c-e) [resp. (f-h)] the sensitivity gain as a function of $\theta_A$ and $\theta_B$ for the different configuration shown in (a) [resp. (b)] (colored regions correspond to $\mathcal{G}^2 \geq 1$). 
Overall, these results show that the mode swapping method allows for a good trade-off between high sensitivity gain for small common phase shift and high bandwidth for large common phase shift, at the expense of generating only a single entangled state.

\subsubsection{Optimal mode-swapping parameters}

\begin{figure}[t!]\includegraphics[width=\columnwidth]{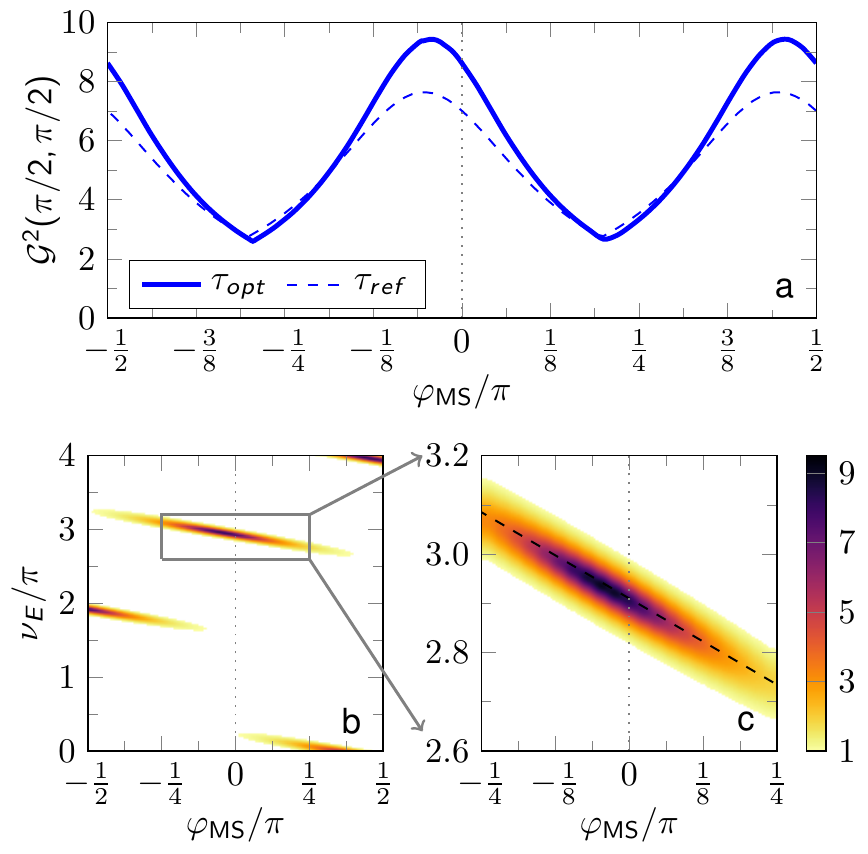}
\caption{Top: Sensitivity gain, $\mathcal{G}^2(\pi/2,\pi/2)$, optimized over $\m$, as a function of $\varphi_{\rm MS}$, for $\tau_E/\tau_{\rm ref}=0.95$ (solid line) and $\tau_E/\tau_{\rm ref}=0.63$ (dashed line). Bottom: $\mathcal{G}^2(\pi/2,\pi/2)$ as a function of $\m$ (the orientation angle of the quantum state after entanglement dynamic), and $\varphi_{\rm MS}$ (the phase of the MS transformation). Panel (c) is a zoom around the optimal parameter values. Here, $\tau_E/\tau_{\rm ref}=0.95$ and $N=100$.}
\label{fig_4}
\end{figure}

The numerical results shown in the Figs.~\ref{fig_2} and \ref{fig_3} have been obtained \new{in the case of a 50-50 beam splitter, i.e. $\vartheta_{\rm MS}=\pi/2$,} for a specific phase value of the MS operation, namely $\varphi_{\rm MS}=\pi/2$. 
We are now interested in the optimization of $\varphi_{\rm MS}$, still keeping $\vartheta_{\rm MS}=\pi/2$. 
In Fig.~\ref{fig_4}(a), we plot $\mathcal{G}^2(\pi/2,\pi/2)$ as a function of $\varphi_{\rm MS}$ for two different squeezing strength $\tau_E$ and for optimized rotations of the squeezing strength, $\m$. 
Here, we observed a $\pi/2$-periodic sinusoidal behavior of the sensitivity gain (predicted analytically in App.~\ref{AppB1}).  
Comparing the two curves, we note that the optimal sensitivity gain is obtained for slightly different optimal phase value $\varphi_{\rm MS}$ which tends, in the regime of low squeezing, to $\varphi_{\rm MS}=-\pi/8$, a result anticipated in the discussion of the analytical model, Sec. \ref{HP}.
 
In Fig.~\ref{fig_4}(b) we show $\mathcal{G}^2(\pi/2,\pi/2)$ (color plot) as a function of the orientation of the spin squeezed state, $\m$, and the phase of the MS, $\varphi_{\rm MS}$, in the specific case $N=100$ and $\tau_E/\tau_{\rm ref}=0.95$ [corresponding to the maximum value of the gain in Fig.~\ref{fig_2}, thin blue line].
This clarifies the dependency of the sensitivity gain on the different parameters of the MePe strategy, 
We note that the orientation of the squeezed state is $4\pi$-dependent and there is a linear dependency between $\nu_{\rm opt}$ and $\varphi_{\rm opt}$ [see the zoom in panel (c)]. 
These results agree with the analytical model, see App. \ref{AppB1} and App. \ref{AppB2}. 
In particular, Fig. \ref{fig_4}(b) should be compared with Fig.~\ref{figSupp_2} in App.~\ref{AppB2}.
Two conclusions can be drawn here: i) the optimal orientation of the squeezed state and the phase of the mode swapping are linearly linked to each other, and ii) the optimal values of $\m$ and $\varphi_{\rm MS}$ are robust against experimental imperfection where phase shift as big as $\pi/8$ still enable large sensitivity gain. 

\begin{figure*}[t!]
\includegraphics[width=\textwidth]{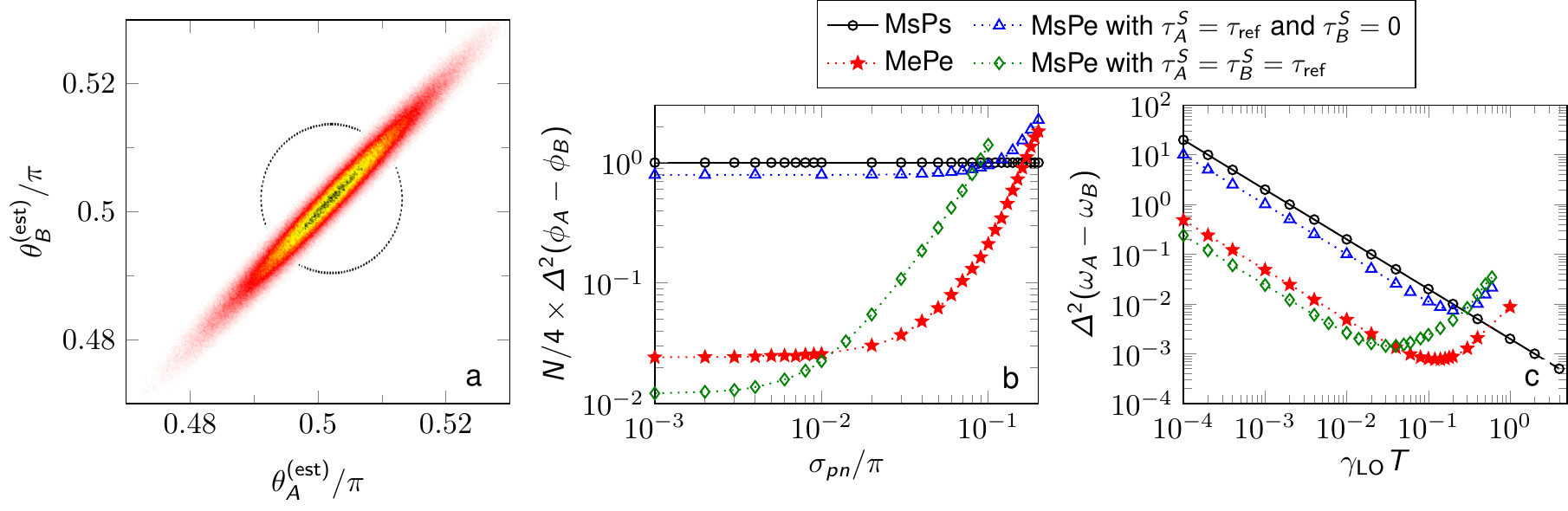}
\caption{(a) Statistical distribution $P(\theta_{A}^{({\rm est})}, \theta_{B}^{({\rm est})})$ of the phase estimators, obtained for the MePe scheme with $\tau_E/\tau_{\rm ref}=0.75$ and $\NT=2000$. 
As a reference, the dotted circles has a radius $\sqrt{4/\NT}$.
(b) Estimation uncertainty $\Delta^2(\phi_A-\phi_B)$, rescaled by the SQL, as a function of the phase noise width $\sigma_{\rm pn}$.
(c) Fractional frequency variance as a function of the interrogation time for $1/f$ noise.
Here we set $\gamma_{\rm LO} \mathcal{T}_{\rm tot}=1$ for simplicity.
In panels (b) and (c) symbols are results of numerical simulations, the solid black line is the SQL and the dotted lines are guides to the eye. 
Stars correspond to the MePe setting with $\tau_E=0.95\tau_{\rm ref}$, while other symbols correspond to mode-separable cases: circles are obtained with two coherent states, triangles with a spin squeezed state ($\tau_A^S=\tau_{\rm ref}$) and a coherent state ($\tau_B^S=0$), and diamonds with two spin-squeezed state ($\tau_A^S=\tau_B^S =\tau_{\rm ref}$). 
In all panels $\NT=2000$.
}
\label{fig_5}
\end{figure*}

\subsection{Examples}
\label{Examples}

\subsubsection{Differential phase estimation}

To illustrate our results, we consider the estimation of the differential phase shift $\phi_A-\phi_B$ in the presence of a common-mode dephasing noise.  
The phase shifts in the two interferometers of Fig.~\ref{fig_1} are $\theta_A = \phi_A+\phi_{\rm pn}$ and $\theta_B=\phi_B+\phi_{\rm pn}$, respectively, where $\phi_{\rm pn}$ is a stochastic phase noise (fluctuating randomly from shot to shot) with a Gaussian distribution centered to zero and with width $\sigma_{\rm pn}$.  
We use the method of moments to estimate $\theta_A$ and $\theta_B$ separately. 
More explicitly, we calculate the joint probability of possible measurement results, $P(N_a, N_b, N_c, N_d\vert \theta_A, \theta_B)$, where $N_j$ is the number of particles at the output port $j=a,b,c,d$ (with $\sum_j N_j = \NT$). 
From this distribution we extract random single shot results. 
We then take $\mu_A = (N_a - N_b)/2$ and $\mu_B = (N_c - N_d)/2$ to infer the expectation values $\mean{\Jz{a,b}}$ and $\mean{\Jz{c,d}}$, respectively. 
The latter quantities are related to the phase shift via $\mean{\Jz{a,b}}=\mean{\Jz{a,b}}_{\rm inp} \cos \theta_{A} + \mean{\Jx{a,b}}_{\rm inp} \sin \theta_{A}$ and $\mean{\Jz{c,d}}=\mean{\Jz{c,d}}_{\rm inp} \cos \theta_{B} + \mean{\Jx{c,d}}_{\rm inp} \sin \theta_{B}$.
Inverting the equations $\mean{\Jz{a,b}}_{\rm out}= \mu_{A}$ and $\mean{\Jz{c,d}}_{\rm out}= \mu_{B}$ gives an estimate $\theta_{A,B}^{({\rm est})}$ of $\theta_{A,B}$, respectively.
In Fig.~\ref{fig_5}(a), we plot the joint distribution $P(\theta_{A}^{({\rm est})}, \theta_{B}^{({\rm est})})$, as obtained from many single-shot measurements, for $\phi_A=\phi_B=\pi/2$, $\sigma_{\rm pn}/\pi=0.01$, $N=2000$ and squeezing parameter $\tau_E/\tau_{\rm ref}=0.75$ [corresponding to the maximum of the red line in Fig.~\ref{fig_2}(a)].
The strong anisotropy of the distribution is due to the correlations between $\theta_A^{({\rm est})}$ and $\theta_B^{({\rm est})}$.
The distribution has the narrowest width along the $\theta_A^{({\rm est})} - \theta_B^{({\rm est})}$ axis, as desired.
As a reference, the black circle in Fig.~\ref{fig_5}(a) has a SQL radius.
The quantity of interest, $\phi_A - \phi_B$, is estimated by taking $\theta_A^{({\rm est})} - \theta_B^{({\rm est})}$.
In Fig.~\ref{fig_5}(b), we plot the uncertainty $\Delta^2 (\phi_A - \phi_B)$ as a function of the noise width $\sigma_{\rm pn}$.
As we see, the uncertainty is well below the SQL (solid black line), up to values of $\sigma_{\rm pn}/\pi \approx 0.2$, in this case.
The increase of $\Delta^2 (\phi_A - \phi_B)$ is directly associated to the finite bandwidth of the spin-squeezed state due to the unavoidable bending in the Bloch sphere.
For comparison, we also plot the sensitivity obtained with a mode-separable strategy with i) two coherent spin states ($\tau_S^A=\tau_S^B=0$, black circles); ii) a coherent state and an optimal spin-squeezed state (
$\tau_S^A/\tau_{\rm ref}=1$ and $\tau_S^B=0$, blue triangles); and iii) two optimal spin-squeezed states ($\tau_S^A=\tau_S^B=\tau_{\rm ref}$, \new{green diamonds}).
The simulations of the differential phase estimation confirm the advantage of the MePe strategy over the MsPe one when using the same resources. 

\subsubsection{Differential frequency estimation}
We consider here relative frequency estimation within an atomic clock scheme.
In this case, we have two atomic ensembles interrogated by the same local oscillator (LO).
The phase shift in the Ramsey interferometer A and B, accumulated during the interrogation time $T$, is 
\be \label{thetaCLOCK}
\theta_{A,B} = \frac{\pi}{2} + \int_T \big(\omega_{\rm LO}(t)- \omega_{A,B} \big) d t
\ee
where $\omega_{A,B}$ is the atomic frequency of the atoms in the interferometer $A$ and $B$, respectively, assumed time independent.
The offset $\pi/2$ in Eq.~(\ref{thetaCLOCK}) places the phase shifts around the optimal working point.  
The accumulated frequency noise due to LO fluctuations is common mode and cancels when evaluating $\phi_A-\phi_B = T (\omega_A-\omega_B)$. 
Within a MsPs strategy, using two independent interferometers with $N$ uncorrelated atoms in total, the fractional frequency variance is \new{given by the standard quantum limit (corresponding to the smallest quantum Cramer-Rao bound for the MsPs strategy~\cite{GessnerPRL2018}),}
\be \label{SQLfreq}
\Delta^2 \Big( \frac{\omega_A-\omega_B}{\omega_0} \Big)_{\rm SQL} = \frac{4}{\omega_0^2 \NT T \mathcal{T}_{\rm tot}}, 
\ee
where $\mathcal{T}_{\rm tot}$ is the total effective interrogation time, given by a multiple of the Ramsey time and $\omega_0$ is a reference value, given by $\omega_0 =(\omega_A+ \omega_B)/2$, for instance.
Using the MePe scheme, we can overcome Eq.~(\ref{SQLfreq}) for relatively short interrogation times.
We find   
\be
\label{ENTfreq}
\Delta^2 \Big( \frac{\omega_A-\omega_B}{\omega_0} \Big) =\frac{4.4}{\omega_0^2 N^{3/2} T \mathcal{T}_{\rm tot}}, 
\ee
for an optimal squeezing time.
Equation (\ref{ENTfreq}) holds for interrogation times $T$ limited by the coherence of the LO, namely when $\theta_A$ and $\theta_B$ remain sufficiently close to the optimal working point $\theta_{A,B}=\pi/2$.
Figure~\ref{fig_5}(c) shows the results of numerical simulations using a LO with $1/f$ power spectral density, as relevant in experiments. 
This can be modelled as a Gaussian phase noise distribution with zero mean and time-dependent variance $\sigma^2_{\rm pn}(T) = (\gamma_{\rm LO} T)^2$, where the parameter $\gamma_{\rm LO}$ sets the coherence time of the laser. 
The states and parameters considered in Fig.~\ref{fig_5}(c) are the same as in Fig.~\ref{fig_5}(b).
In particular, the black circles are obtained for the \new{MsPs} setting with two coherent spin state of $N/2$ particles each.  
The \new{red dotted} line is Eq.~(\ref{SQLfreq}).
The \new{red stars} show the results of our MS scheme.
For sufficiently short $T$, our scheme follows Eq. (\ref{ENTfreq}) and provides a substantial advantage with respect to the SQL: in other words we can reach the same absolute sensitivity as the fully separable case but with a $O(1/\sqrt{\NT})$ decrease of averaging time, which can be an important advantage for some applications.  
The figure assumes that the atomic coherence lifetime is longer than the LO coherence time. 
\new{In the opposite limit, the bending of the variance happens at interrogation shorter times, in the regime of validity of Eqs.~(\ref{SQLfreq}) and~(\ref{ENTfreq}).
In this case,}
the mode-entangled strategy can improve the overall absolute fractional frequency variance by a factor $\sqrt{\NT}$.
Differently from LO stabilization in atomic clock, differential frequency estimation does not suffer the Dick effect associated to dead times \cite{SchulteNATCOMM2020}, even when using squeezed states. 

\section{Discussion}

\subsection{\new{Possible generalizations}}

In the manuscript, we have focused on the use of \new{spin}-squeezed states generated via one-axis twisting dynamics. 
However, the protocol can be immediately extended to various squeezing procedures.
In fact, within a Holstein-Primakoff transformation, we have seen that the differential sensitivity is directly related to the quadrature variance $\Delta^2(x_b+x_c)$, see Eq.~(\ref{HPapprox}): any state that realizes $\Delta^2(x_b+x_c)<1$ can be used to reduce the differential uncertainty below the SQL. 
The case of cavity squeezing and two-axis counter-twisting are briefly discussed in App.~\ref{AppD}. 
An interesting case is that of the two-mode squeezed vacuum, $\ket{\psi} = e^{-i r (\hat{b}^\dag \hat{c}^\dag + \hat{b}\hat{c})} \ket{0}_b \ket{0}_c$, that has been realized in atomic system via spin-mixing dynamics in spinor Bose-Einstein condensates~\cite{HamleyNATPHYS2012, GrossNATURE2011, PeiseNATCOMM2015}.
This state is characterized by $\Delta^2(\hat{x}_b+\hat{x}_c) = 2e^{-2r}$~\cite{BarnettBOOK} and thus satisfies $\Delta^2(\hat{x}_b+\hat{x}_c)<1$ without requiring the MS operation, provided that $r>(\log 2)/2 \approx 0.35$. 
Interestingly, the inequality $r>(\log 2)/2$ is also the condition for the two-mode squeezed vacuum state to fulfills the Einstein-Podolsky-Rosen (EPR) criterion~\cite{ReidPRA1989, ReeidRMP2009, OuPRL1992, PeiseNATCOMM2015}. 
Recently, EPR entanglement has been proposed as a resource for optical differential homodyne measurements~\cite{MaNATPHYS2017, SudbeckNATPHOT2020}.
In contrast, within our MS scheme, any single-mode squeezed states with positive value of the squeezing parameter, $r>0$ (or, equivalently, any one-axis twisted state with arbitrary small spin-squeezing), can realize a sub-SQL sensitivity.
As single mode squeezed states do not fulfill the EPR criterion,
we thus conclude that EPR entanglement is a useful resource for differential phase estimation with sub-SQL sensitivity (in agreement with Refs.~\cite{MaNATPHYS2017, SudbeckNATPHOT2020}), although it is not a strictly necessary resource. 

Besides differential phase estimation, our scheme can also be extended to the sub-SQL estimation of an arbitrary linear combination of the two phase shifts, $\nu_A \theta_A + \nu_B \theta_B$, where $\nu_{A,B}$ are arbitrary real numbers \cite{MalitestaARXIV}.
This extension simply requires to tune the MS operation (see Appendix for details). 

Finally, it is worth emphasizing that the MePe protocol studied in this manuscript can be combined with existing protocols to extend the squeezed state sensitivity bandwidth, for instance using adaptive schemes~\cite{PezzePRL2020, PezzePRX2021}, nonlinear~\cite{KaubrueggerPRX2021, MarciniakNATURE2022} or non-demolition \cite{BorreggardPRL2013, KohlhaasPRX2015, BowdenPRX2020} measurements.


\new{\subsection{Possible experimental implementation with Bose-Einstein condensates}}
\new{
On the overall, the input four-mode entangled state proposed in Fig.~\ref{fig_1}a could be experimentally generated with a suitable combination of Raman and Bragg diffraction, on ground fountain experiment~\cite{Asenbaum20,Janvier22} or in space environment~\cite{Gaaloul22} for example. 
Based on recent experimental development and new methods in the field, we highlight a possible experimental sequence with Bose-Einstein condensate (BEC).
}

\new{Considering the mode $a\equiv |\uparrow,0\rangle$, $b\equiv |\downarrow,2\hbar k\rangle$, $c\equiv |\downarrow,0\rangle$ and $d\equiv |\uparrow,2\hbar k\rangle$ with $2\hbar k$ the momentum transfer, the four-mode state can be engineer as follow:}
\new{(i) a BEC, initially in mode $a$ is diffracted onto a quantum superposition of same internal state with different external momentum, e.g. coupling mode $a$ and $d$, though a first Bragg pulse generating the state $|\Psi_1\rangle$ [yellow box].}
\new{(ii) a Raman laser couple mode $a$ and $b$ and entanglement dynamic is generated with the method of the Delta-kick Squeezing~\cite{PRLCorgier21}.
This method allow us to access various value of squeezing strength, $\tau_E$, through the control of atomic collision.
The orientation of the quantum state, $\nu_E$, can then be manipulated through the phase of the laser transition creating the state $|\Psi_2\rangle$  [green box].}
\new{(iii) The MS beam-splitter can be engineered with a second Bragg pulse coupling mode $b$ and $c$.
The parameter $\varphi_{\rm MS}$ is controlled by the phase of the laser, creating the entangled four-mode state $|\Psi_{\rm inp}\rangle$ [blue box].
Bringing modes $a$ and $b$ and modes $c$ and $d$ to spatially overlap, with possible momentum transfer~\cite{AndersPRL2021}, the state $|\Psi_{\rm inp}\rangle$ can be used for a differential measurement.}\\

\section{Conclusions}

We have demonstrated a protocol for differential phase measurements with sensitivity enhanced by spin squeezing. 
A single two-mode spin-squeezed state is optimally manipulated via a mode-swapping operation that, although consisting essentially in a simple beam splitter, is enough to generate useful mode entanglement. 
Our mode-entangled and particle-entangled protocol overcomes the SQL and benefits from the cancellation of common-mode noise. 

Our scheme represents a practical example of distributed quantum sensing \cite{ProctorPRL2018, GePRL2018, GuoNATPHYS2020, XiaPRL2020, LiuNATPHOT2021} where mode entanglement plays a crucial role.
In particular, the corresponding particle-entangled mode-separable protocol using a single spin-squeezed state and a coherent state in independent interferometers cannot overcome a sensitivity scaling $1/N$, with only a unit gain factor over the SQL. 
In contrast, our scheme achieves, with the same resources, a scaling $1/N^{3/2}$ with one-axis-twisted states and may even reach the Heisenberg scaling $1/N^2$ with highly squeezed states.

To conclude, we notice that, after the completion of this work, a related sub-SQL differential scheme appeared~\cite{MaliaARXIV}. 
There, an atomic distributed quantum sensing protocol is demonstrated experimentally by squeezing the total spin of the two interferometers, e.g. $\hat{J}_x^{a,b}+\hat{J}_x^{c,d}$, see Eq. (\ref{DthetaABpi2}), via atom-light quantum non-demolition measurements in a cavity.
In our proposal, we squeeze only one collective spin, e.g. $\hat{J}_x^{a,b}$, and then transfer the squeezing to the sum $\hat{J}_x^{a,b}+\hat{J}_x^{c,d}$ via mode swapping.\\
\\*
{\bf Acknowledgments.}
R.C. thanks the Paris Observatory Scientific Council 
and was funded by "PSL fellowship at Paris Observatory" program.
We acknowledge financial support from the QuantEra project SQUEIS. 
We also acknowledge financial support from the European Union’s Horizon 2020 research and innovation programme—Qombs Project, FET Flagship on Quantum Technologies Grant No. 820419.

\clearpage
\appendix
\onecolumngrid

\section{Details on the analytical approach and approximations}
\label{AppA}

\subsection{Interferometer operations} 
\label{AppA1}

\paragraph{} Using the Stirling formula $N! \sim \sqrt{2\pi N} N^N e^{-N}$, valid for $N \gg 1$, and absorbing the factor $\sqrt{2\pi N}$ in the normalization of the state, the state Eq. (\ref{psi1}) writes
\beq \label{step1}
\ket{\psi_1} &\approx& \sum_{m=-N/2}^{N/2}  \frac{N^{N/2} e^{-N/2}}{\sqrt{(N/2-m)!  (N/2+m)!}} \ket{N/2+m, 0, 0, N/2-m} \nonumber \\
&\approx& \sum_{n_a=0}^{+\infty}  \frac{\alpha_a^{n_a} e^{-\vert \alpha_a \vert^2/2}}{\sqrt{n_a!}}
\sum_{n_b=0}^{+\infty}  \frac{\alpha_b^{n_b} e^{-\vert \alpha_b \vert^2/2}}{\sqrt{n_b!}} \delta_{n_a+n_b,N}\,\ket{n_a, 0, 0, n_b}.
\eeq
The delta function $\delta_{n_a+n_b,N}$ expresses the fixed total number of particles. 
Neglecting this term allows for coherences between total number of particles which are, in turns, eliminated when measuring the total number of particles in output.  
Neglecting the delta function, the state $\ket{\psi_1}$ writes as $\ket{\psi_1} = \ket{C(\alpha_a)}\ket{0}_b\ket{0}_c \ket{C(\alpha_d)}_d$, where $\alpha_a=\alpha_d=\sqrt{N/2}$. 
As a generalization, the single-mode coherent states $\ket{C(\alpha_a)}$ and $\ket{C(\alpha_d)}$ with $\alpha_a=|\alpha_a|e^{i\phi_{a,0}}$ and $\alpha_d=|\alpha_d|e^{i\phi_{d,0}}$, respectively, will be used. 
For $|\alpha_a|=|\alpha_d|=\sqrt{N/2}$ and $\phi_{a,0}=\phi_{d,0}=0$, we recover the discussion in the main text.

\paragraph{} The second step of the protocol is given by the one-axis twisting evolution $\exp\{-i\tau_E(\Jx{a,b})^2\}$ followed by the rotation $\exp\{-i\m \Jz{a,b}\}$.
Within the Holstein-Primakoff transformation, we have $\hat{a}\sim e^{i\phi_{a,0}}\sqrt{N/2}$, and thus
\begin{align}
\label{psi_s}
    &\exp\{-i\tau_E(\Jx{a,b})^2\}\ket{C(\alpha_a)}_a\ket{0}_b\ket{0}_c \ket{C(\alpha_d)}_d
    = \exp\{-i\tau_E/4\left(\hat{a}^{\dagger}\hat{b}+\hat{a}\hat{b}^{\dagger}\right)^2\}\ket{C(\alpha_a)}_a\ket{0}_b\ket{0}_c \ket{C(\alpha_d)}_d\nonumber\\
    \approx&\exp\{-ie^{i\phi_{a,0}} N\tau_E/8\left(\hat{b}+\hat{b}^{\dagger}\right)^2\}\ket{C(\alpha_a)}_a\ket{0}_b\ket{0}_c \ket{C(\alpha_d)}_d\nonumber\\
    \approx&\exp\{e^{i\left(\phi_{a,0}-\pi/2\right)}N\tau_E/8\left(\hat{b}^2+{\hat{b}^{\dagger^2}}\right)\}\exp\{-iN\tau/8\left(2\hat{b}^{\dagger}\hat{b}+1\right)\}\ket{C(\alpha_a)}_a\ket{0}_b\ket{0}_c \ket{C(\alpha_d)}_d\\
    =&\exp\{e^{i\left(\phi_{a,0}-\pi/2\right)}N\tau_E/8\left(\hat{b}^2+{\hat{b}^{\dagger^2}}\right)\}\ket{C(\alpha_a)}_a\ket{0}_b\ket{0}_c \ket{C(\alpha_d)}_d\nonumber \\
    =& \ket{C(\alpha_a)}_a \ket{S(\xi_0)}_b \ket{0}_c \ket{C(\alpha_d)}_d,\nonumber
\end{align}
where the state $\ket{S(\xi_0)}_b=\hat{S}_b(\xi_0)\ket{0}_b=\exp\{(\xi_0^*\hat{b}^2-\xi_0{\hat{b}^{\dagger^2}})/2\}\ket{0}_b$ is a single-mode squeezed-vacuum state, with squeezing parameter $\xi_0=e^{i\left(\pi/2-\phi_{a,0}\right)}N\tau_E/4$. In the derivation of the moment matrix, we will use $\xi_0=re^{i\varphi_0}$ as a general squeezing parameter. 
The case discussed in the main text corresponds to the specific choice $r=N\tau_E/4$ and $\varphi_0=\pi/2-\phi_{a,0}=\pi/2$. After squeezing, the rotation $\exp\{-i\m\Jz{a,b}\} = \exp\{-i(\m/2)\hat{a}^{\dagger}\hat{a}\}\exp\{i(\m/2)\hat{b}^{\dagger}\hat{b}\}$ is applied to modes $a$ and $b$, thus affecting the phases $\phi_{a,0}$ and $\varphi_0$ that become $\phi_a=\phi_{a,0}-\m/2$ and $\varphi = \varphi_0+\m$, respectively. At this stage, we have
\begin{align}
    \ket{\psi_2}=\ket{C(\alpha_a e^{-i\m/2})}_a \ket{S(\xi)}_b \ket{0}_c \ket{C(\alpha_d)}_d,\nonumber
\end{align}
where $\xi=re^{i\varphi}$.

\paragraph{} A mode-swapping operation is then applied between modes $b$ and $c$. 
Here, we assume a general unitary mode-swapping transformation:
\begin{align}\label{general_U_matrix}
    \begin{pmatrix}\hat{U}_{\rm MS}^{\dagger}\hat{b}\hat{U}_{\rm MS}\\\hat{U}_{\rm MS}^{\dagger}\hat{c}\hat{U}_{\rm MS}\end{pmatrix}=&\begin{pmatrix}|u_{bb}|e^{i\delta_{bb}}&-|u_{cb}|e^{-i\delta_{bc}}\\|u_{cb}|e^{i\delta_{cb}}&|u_{bb}|e^{-i\delta_{cc}}\end{pmatrix}\begin{pmatrix}\hat{b}\\\hat{c}\end{pmatrix},
\end{align}
with $\delta_{cc}=\delta_{bb}+\delta_{bc}-\delta_{cb}$. The more specific case of Eq. \makeref{U_matrix} can be recovered by setting $\delta_{bb}=\delta_{cc}=0$,  which implies $\delta_{bc}=\delta_{cb}$.  

\paragraph{} As a final stage, the two interferometers $\exp\{-i\theta_A\hat{J}_{y,\{a,b\}}\}$ and $\exp\{-i\theta_B\hat{J}_{y,\{c,d\}}\}$ are used to encode the two phase shifts $\theta_A$ and $\theta_B$ to be estimated, implementing the transformation:
$\ket{\psi_{\rm out}}=\exp\{-i\theta_A\Jy{a,b}\}\exp\{-i\theta_B\Jy{c,d}\}\ket{\psi_{\rm inp}}$.

\subsection{Phase sensitivity equation}
\label{AppA2}

The uncertainty in the estimation of a generic linear combination $\vect{v} \cdot \vect{\theta} = \nu_A \theta_A + \nu_B \theta_B$ using the method of moments \cite{GessnerNATCOMM2020, MalitestaARXIV} is
\begin{align} \label{arbsens}
\Delta^2(v_A\theta_A+v_B\theta_B)&=v_A^2\frac{\Delta^2\Jz{a,b}\vert_{\rm out}}{\left[\frac{\partial}{\partial\theta_A}\bra{\psi_{\rm out}}\Jz{a,b}\ket{\psi_{\rm out}}\right]^2}+v_B^2\frac{\Delta^2\Jz{c,d}\vert_{\rm out}}{\left[\frac{\partial}{\partial\theta_B}\bra{\psi_{\rm out}}\Jz{c,d}\ket{\psi_{\rm out}}\right]^2}\nonumber \\
&+ 2v_Av_B\frac{\textrm{cov}\left(\Jz{a,b},\Jz{c,d}\right)_{\rm out}}{\frac{\partial}{\partial\theta_A}\bra{\psi_{\rm out}}\Jz{a,b}\ket{\psi_{\rm out}}\frac{\partial}{\partial\theta_B}\bra{\psi_{\rm out}}\Jz{c,d}\ket{\psi_{\rm out}}},
\end{align}
where $\nu_{A,B}$ are arbitrary real numbers
In the case $\vect{\nu}=(1,-1)$ Eq. (\ref{arbsens}) reduces to Eq.~(\ref{DthetaAB}).
Following Ref. \cite{MalitestaARXIV} and using the approximated state Eq. (\ref{psinpHP}) as input, we obtain 

\begin{align}\label{general_sensitivity}
\Delta^2(\vect{v}\cdot\vect{\theta})=&\left(e^{2r}-1\right)\left(\frac{|\alpha_a||u_{bb}|\sin{\chi_A}}{|\alpha_a|^2-|u_{bb}|^2s^2}v_A-\frac{|\alpha_d||u_{cb}|\sin{\chi_B}}{|\alpha_d|^2-|u_{cb}|^2s^2}v_B\right)^2\nonumber\\
-&\left(1-e^{-2r}\right)\left(\frac{|\alpha_a||u_{bb}|\cos{\chi_A}}{|\alpha_a|^2-|u_{bb}|^2s^2}v_A-\frac{|\alpha_d||u_{cb}|\cos{\chi_B}}{|\alpha_d|^2-|u_{cb}|^2s^2}v_B\right)^2\nonumber\\
+&\frac{|\alpha_a|^2+|u_{bb}|^2s^2}{\left(|\alpha_a|^2-|u_{bb}|^2s^2\right)^2}v_A^2+\frac{|\alpha_d|^2+|u_{cb}|^2s^2}{\left(|\alpha_d|^2-|u_{cb}|^2s^2\right)^2}v_B^2+\mathcal{Q}(\cot{\theta_A},\cot{\theta_B}),
\end{align}
where we have set $\sinh{r}\equiv s$, $\cosh{r}\equiv c$,
\begin{align}\label{general_sensitivity_Q}
    \mathcal{Q}(\cot{\theta_A},\cot{\theta_B})=&\cot^2{\theta_A}\frac{|\alpha_a|^2+|u_{bb}|^2s^2}{\left(|\alpha_a|^2-|u_{bb}|^2s^2\right)^2}v_A^2+\cot^2{\theta_B}\frac{|\alpha_d|^2+|u_{cb}|^2s^2}{\left(|\alpha_d|^2-|u_{cb}|^2s^2\right)^2}v_B^2\nonumber\\
    &+\left(2c^2-1\right)s^2\left(\cot{\theta_A}\frac{|u_{bb}|^2}{|\alpha_a|^2-|u_{bb}|^2s^2}v_A+\cot{\theta_B}\frac{|u_{cb}|^2}{|\alpha_d|^2-|u_{cb}|^2s^2}v_B\right)^2,
\end{align}
and 
\begin{align}
    \chi_A=&\phi_a-\frac{\varphi}{2}-\delta_{bb}=\phi_{a,0}-\frac{\varphi_0}{2}-\m-\delta_{bb},\label{chiA}\\
    \chi_B=&\phi_d-\frac{\varphi}{2}-\delta_{cb}=\phi_{d,0}-\frac{\varphi_0}{2}-\frac{\m}{2}-\delta_{cb}\label{chiB}
\end{align}
(notice that $\phi_d=\phi_{d,0}$).

\subsection{Relation between phase sensitivity and quadrature variance}
\label{AppA3}

At the optimal working point $\theta_A=\theta_B=\pi/2$, Eq. (\ref{arbsens}) writes 
\begin{align}
    \Delta^2(v_A\theta_A+v_B\theta_B)\big\vert_{\theta_A=\theta_B=\pi/2}=& \Delta^2\left[ v_A \frac{\Jx{a,b}}{\bra{\psi_{\rm inp}}\Jz{a,b}\ket{\psi_{\rm inp}}} + v_B\frac{\Jx{c,d}}{\bra{\psi_{\rm inp}}\Jz{c,d}\ket{\psi_{\rm inp}}}\right]_{\rm inp},
\end{align}
where the variance is now evaluated on state $\ket{\psi_{\rm inp}}$. 
By taking $v_A=1$, $v_B=-1$ and further working out the right-hand side and assuming $|\alpha_a|=|\alpha_d|=\sqrt{N/2}$, we find
\begin{align}
    \Delta^2(\theta_A-\theta_B)\big\vert_{\theta_A=\theta_B=\pi/2}=& \Delta^2\left[\frac{\hat{a}^{\dagger}\hat{b}+\hat{a}\hat{b}^{\dagger}}{N/2-\bra{\psi_{\rm inp}}\hat{b}^{\dagger}\hat{b}\ket{\psi_{\rm inp}}}+\frac{\hat{c}^{\dagger}\hat{d}+\hat{c}\hat{d}^{\dagger}}{N/2-\bra{\psi_{\rm inp}}\hat{c}^{\dagger}\hat{c}\ket{\psi_{\rm inp}}}\right]_{\rm inp}.
\end{align}
As a last assumption, we take $N$ to be sufficiently large to apply the Holstein-Primakoff transformation: as explained in the main text, this amounts to taking $\hat{a}\sim e^{i\phi_a}\sqrt{N/2}$ and $\hat{d}\sim e^{i\phi_d}\sqrt{N/2}$. Being in the regime of large $N$ will also allow us to neglect the two expectation values $\bra{\psi_{\rm inp}}\hat{b}^{\dagger}\hat{b}\ket{\psi_{\rm inp}}$ and $\bra{\psi_{\rm inp}}\hat{c}^{\dagger}\hat{c}\ket{\psi_{\rm inp}}$ found in the denominators. These two terms represent the average number of particles in modes $b$ and $c$, respectively, after mode-swapping, so that each of them is smaller than $\bar{n}_s$ and can be safely neglected if $N/2\gg\bar{n}_s$. Overall, we find
\begin{align}
    \Delta^2(\theta_A-\theta_B)\big\vert_{\theta_A=\theta_B=\pi/2}\approx&\frac{\Delta^2\left[e^{-i\phi_a}\hat{b}+e^{i\phi_a}\hat{b}^{\dagger}+e^{-i\phi_d}\hat{c}+e^{i\phi_d}\hat{c}^{\dagger}\right]}{N/2}
    =4\frac{\Delta^2\left(\hat{x}_b +\hat{x}_c \right)}{N},\label{var_squeezed}
\end{align}
where $\hat{x}_b=(e^{-i\phi_a}\hat{b}+e^{i\phi_a}\hat{b}^{\dagger})/\sqrt{2}$ and $\hat{x}_c=(e^{-i\phi_d}\hat{c}+e^{i\phi_d}\hat{c}^{\dagger})/\sqrt{2}$. We and thus recover Eq. (\ref{HPapprox}).
Next, we evaluate $\Delta^2\left(\hat{x}_b +\hat{x}_c \right)$ on the mode-swapped squeezed state $\hat{U}_{\rm MS}\ket{S(\xi)}_b\ket{0}_c$:
\be
\Delta^2\left(\hat{x}_b +\hat{x}_c \right)=\frac{4}{N}\mathbin{_b\bra{S(\xi)}}\mathbin{_c\bra{0}}\hat{U}^{\dagger}_{\rm MS}\left(\hat{x}_b +\hat{x}_c \right)^2\hat{U}_{\rm MS}\ket{S(\xi)}_b\ket{0}_c,
\ee
and minimize the uncertainty using the MS transformation $\hat{U}_{\rm MS}$.
We calculate
\begin{align}
&\mathbin{_b\bra{S(\xi)}}\mathbin{_c\bra{0}}\hat{U}^{\dagger}_{\rm MS}\left[\hat{x}_b(\lambda) +\hat{x}_c(\lambda) \right]^2\hat{U}_{\rm MS}\ket{S(\xi)}_b\ket{0}_c= \frac{1}{2}\left(e^{2r}-1\right)\left[|u_{bb}|\sin{\left(\lambda+\chi_A\right)}+|u_{cb}|\sin{\left(\lambda+\chi_B\right)}\right]^2\nonumber\\
&-\frac{1}{2}\left(1-e^{-2r}\right)\left[|u_{bb}|\cos{\left(\lambda+\chi_A\right)}+|u_{cb}|\cos{\left(\lambda+\chi_B\right)}\right]^2+1,
\end{align}
where $\hat{x}_b(\lambda)=(e^{-i\left(\lambda+\phi_a\right)}\hat{b}+e^{i\left(\lambda+\phi_a\right)}\hat{b}^{\dagger})/\sqrt{2}$ and $\hat{x}_c(\lambda)=(e^{-i\left(\lambda+\phi_d\right)}\hat{c}+e^{i\left(\lambda+\phi_d\right)}\hat{c}^{\dagger})/\sqrt{2}$ are generalized quadrature operators. 
Note the similarity of this equation with Eq. \makeref{specific_sensitivity} (next section), for $\lambda=0$ and in the limit of large $N$, which is consistent with our previous discussion. Applying no mode-swapping to the state $\ket{S(\xi)}_b\ket{0}_c$ would imply
\begin{align}
\mathbin{_b\bra{S(\xi)}}\mathbin{_c\bra{0}}\left[\hat{x}_b(\lambda) +\hat{x}_c(\lambda) \right]^2\ket{S(\xi)}_b\ket{0}_c= \frac{1}{2}\left(e^{2r}+1\right)\sin^2{\left(\lambda+\chi_A\right)}+\frac{1}{2}\left(e^{-2r}+1\right)\cos^2{\left(\lambda+\chi_A\right)}.
\end{align}
Even if the state $\ket{S(\xi)}_b$ was optimized through a rotation in the $\hat{x}_b$-$\hat{p}_b$ quadrature plane, setting $\chi_A=\phi_a-\varphi/2=0$, that would correspond, for $\lambda=0$, to an estimation uncertainty $\Delta^2(\theta_A-\theta_B)\big\vert_{\theta_A=\theta_B=\pi/2}\approx 2e^{-2r}/N+2/N$, implying a sensitivity gain $\mathcal{G}^2=2$, at best, in the limit $r\to +\infty$. Hence, the high sensitivity gain typical of one-parameter estimation cannot be achieved here by a simple rotation in the $\hat{x}_b$-$\hat{p}_b$ plane. To that aim, the effect of that rotation needs to be combined with a mode-swapping operation. To understand how mode-swapping manages to enhance the sensitivity of the differential measurement, consider the chain of equalities:
\begin{align}
&\mathbin{_b\bra{S(\xi^{\rm opt})}}\mathbin{_c\bra{0}}\hat{x}_b^2(\lambda-\delta_{bb})\ket{S(\xi^{\rm opt})}_b\ket{0}_c=\mathbin{_b\bra{S(\xi^{\rm opt})}}\mathbin{_c\bra{0}}\hat{U}_{\rm MS}^{\dagger}\left[\hat{U}_{\rm MS}\hat{x}_b^2(\lambda-\delta_{bb})\hat{U}_{\rm MS}^{\dagger}\right]\hat{U}_{\rm MS}\ket{S(\xi^{\rm opt})}_b\ket{0}_c\\
&=\mathbin{_b\bra{S(\xi^{\rm opt})}}\mathbin{_c\bra{0}}\hat{U}_{\rm MS}^{\dagger}\Bigg[|u_{bb}|\frac{e^{-i\left(\lambda+\phi_a\right)}\hat{b}+e^{i\left(\lambda+\phi_a\right)}\hat{b}^{\dagger}}{\sqrt{2}}\nonumber\\
&\qquad\qquad\qquad\qquad\quad+|u_{cb}|\frac{e^{-i\left(\lambda-\delta_{bb}+\phi_a+\delta_{cb}\right)}\hat{c}+e^{i\left(\lambda-\delta_{bb}+\phi_a+\delta_{cb}\right)}\hat{c}^{\dagger}}{\sqrt{2}}\Bigg]^2\hat{U}_{\rm MS}\ket{S(\xi^{\rm opt})}_b\ket{0}_c\label{interm}\\
&=\frac{1}{2}\mathbin{_b\bra{S(\xi^{\rm opt})}}\mathbin{_c\bra{0}}\left(\hat{U}^{\rm opt}_{\rm MS}\right)^{\dagger}\left[\hat{x}_b(\lambda) +\hat{x}_c(\lambda) \right]^2\hat{U}_{\rm MS}^{\rm opt}\ket{S(\xi^{\rm opt})}_b\ket{0}_c.\label{final}
\end{align}
As shown in the next section, optimization of both the state $\ket{S(\xi)}_b$ and the mode-swapping parameters, in the large $N$ regime, requires $\chi_A=0$, $\chi_B=0$ and $|u_{bb}|=|u_{cb}|=1/\sqrt{2}$. Application of these optimal conditions leads to Eq. \makeref{final}, in which the differential quadrature operator $\hat{x}_b({\lambda})+\hat{x}_c({\lambda})$ already found above is recovered. This proves that, if evaluated on the mode-swapped state $\hat{U}_{\rm MS}^{\rm opt}\ket{S(\xi^{\rm opt})}_b\ket{0}_c$, the variance of $\hat{x}_b({\lambda})+\hat{x}_c({\lambda})$ reduces to that of $\hat{x}_b({\lambda}-\delta_{bb})$ on $\ket{S(\xi^{\rm opt})}_b$. Thus, equivalent results to those obtained in one-parameter estimation can be reproduced in a differential measurement using a single squeezed, mode-swapped state. More specifically, one gets
\begin{align}
    \mathbin{_b\bra{S(\xi^{\rm opt})}}\mathbin{_c\bra{0}}\left(\hat{U}^{\rm opt}_{\rm MS}\right)^{\dagger}\left[\hat{x}_b({\lambda}) +\hat{x}_c({\lambda}) \right]^2\hat{U}_{\rm MS}^{\rm opt}\ket{S(\xi^{\rm opt})}_b\ket{0}_c=e^{2r}\sin^2{\lambda}+e^{-2r}\cos^2{\lambda}.
\end{align}
For $\lambda=0$ and using Eq. \makeref{var_squeezed},  $\Delta^2(\theta_A-\theta_B)\big\vert_{\theta_A=\theta_B=\pi/2}\approx 4e^{-2r}/N$, with a sensitivity gain now given by $\mathcal{G}^2=e^{2r}$. Of course, the gain is not expected to increase indefinitely with squeezing and, as shown in the next section, a more accurate calculation predicts the existence of a maximum achievable gain: $\mathcal{G}^2_{\rm max}=\sqrt{N}$.

\section{Optimization}
\label{AppB}

\subsection{Simultaneous optimization of the $\hat{J}_z$ rotation and the mode-swapping parameters}
\label{AppB1}

Our goal here is to determine the rotation angle $\nu_E$ in $\exp\{-i\m \Jz{a,b}\}$ and the mode-swapping parameters, Eq. \makeref{general_U_matrix}, that minimize the uncertainty of our estimation method, Eq. \makeref{general_sensitivity}. 
In particular, we consider the case $v_A=1$, $v_B=-1$ (differential measurement case) and $|\alpha_a|^2=|\alpha_d|^2=N/2$.
As a first step, we minimize that function with respect to the phases $\chi_A$ and $\chi_B$ defined in Eqs. \makeref{chiA} and \makeref{chiB}, respectively. Given that $Q(\cot{\theta_A},\cot{\theta_B})$ is independent of these two parameters, we just focus on 
\begin{align}
\Delta^2(\theta_A-\theta_B)\big\vert_{\theta_A=\theta_B=\pi/2}&= N\left(e^{2r}-1\right)\left(\frac{|u_{bb}|\sin{\chi_A}}{N-|u_{bb}|^2\bar{n}_s}+\frac{|u_{cb}|\sin{\chi_B}}{N-|u_{cb}|^2\bar{n}_s}\right)^2\nonumber\\
&-N\left(1-e^{-2r}\right)\left(\frac{|u_{bb}|\cos{\chi_A}}{N-|u_{bb}|^2\bar{n}_s}+\frac{|u_{cb}|\cos{\chi_B}}{N-|u_{cb}|^2\bar{n}_s}\right)^2\nonumber\\
&+\frac{N+|u_{bb}|^2\bar{n}_s}{\left(N-|u_{bb}|^2\bar{n}_s\right)^2}+\frac{N+|u_{cb}|^2\bar{n}_s}{\left(N-|u_{cb}|^2\bar{n}_s\right)^2},\label{specific_sensitivity}
\end{align}
where we have used $\bar{n}_s$ to represent the average number of particles in mode $b$ after squeezing and before the mode-swapping stage: $\bar{n}_s=\sinh^2{r}=\sinh^2{\left(N\tau_E/4\right)}$.
We notice that, by choosing $\sin{\chi_A}=\sin{\chi_B}=0$ (which implies $\cos{\chi_A}=\pm 1$ and $\cos{\chi_B}=\pm 1$), we minimize the first squared term in Eq. \makeref{specific_sensitivity} and \textit{simultaneously} maximize the second one by a suitable choice of the sign of the two cosines. This yields the smallest possible value of the uncertainty, or the \textit{absolute} minimum of Eq. \makeref{specific_sensitivity} with respect to $\chi_A$ and $\chi_B$. In order to maximize the second squared term in that equation, the sum of two terms of equal sign is clearly needed between round brackets. In the regime $N \gg \bar{n}_s$, this requires $\cos{\chi_A}=\cos{\chi_B}=\pm 1$, which, by virtue of Eq. \makeref{chiA} and \makeref{chiB}, translates to 
\begin{equation}\label{conditions}
    \begin{split}
    &\phi_{a,\rm 0}-\frac{\varphi_{\rm 0}}{2}-\m-\delta_{bb}=k\pi+2m\pi,\\
    &\phi_{d,\rm 0}-\frac{\varphi_{\rm 0}}{2}-\frac{\m}{2}-\delta_{cb}=k\pi,
    \end{split}
    \qquad k,m = 0, \pm 1, \dots.
\end{equation}
Moreover, since $\phi_{a,0}=\phi_{d,0}=0$, $\varphi_0=\pi/2$ and $\delta_{bb}=0$ in the actual interferometer scheme, we end up with
\begin{equation}\label{ab_minimum_alpha_delta}
    \begin{split}
    &\frac{\pi}{4}+\m=k\pi+2m\pi,\\
    &\frac{\pi}{4}+\frac{\m}{2}+\delta_{cb}=k\pi,
    \end{split}
    \qquad k,m = 0, \pm 1, \dots,
\end{equation}
which gives 
\begin{equation}
    \begin{split}
        &\delta_{cb} = (4k-1)\frac{\pi}{8}-m\pi,\\
        &\m = (4k-1)\frac{\pi}{4}+2m\pi,
    \end{split}
     \quad k,m = 0, \pm 1, \dots,
\quad {\rm or}
\quad \begin{split}
        &\delta_{cb} = -\frac{\pi}{8}+l\frac{\pi}{2},\\
        &\m = 2\delta_{cb}+4m\pi,
    \end{split}
     \quad l,m = 0, \pm 1, \dots.\label{ab_minimum_alpha_delta_2}
\end{equation}
The second parametrization of the solutions clearly reveals the $4\pi$ periodicity of the sensitivity as function of $\m$; this is consequence of two simple facts: the $2\pi$ periodicity of the sensitivity as function of $\chi_A$ and $\chi_B$, Eq. \makeref{specific_sensitivity}, and the fact that a $4\pi$ rotation in $\m$ is needed to get a full $2\pi$ variation in $\chi_B$, Eq. \makeref{chiB}. By varying $l$ and $m$, we get all the pairs $(\m, \delta_{cb})$ that minimize the function in Eq. \makeref{specific_sensitivity} and, hence, the total uncertainty of the differential measurement.

Making use of the optimal values of the parameters, from Eq. \makeref{specific_sensitivity} one gets
\begin{align}\label{3}
\Delta^2(\theta_A-\theta_B)\big\vert_{\theta_A=\theta_B=\pi/2}=-2\frac{1-e^{-2r}}{N}\left(|u_{bb}|+|u_{cb}|\right)^2+\frac{4}{N}+\frac{4\bar{n}_s}{N^2}, \quad \bar{n}_s\ll N.
\end{align}
As a function of the amplitudes $|u_{bb}|$ and $|u_{cb}|$, this is clearly minimized when the sum $|u_{bb}|+|u_{cb}|$ attains its maximum value under the constraint $|u_{bb}|^2+|u_{cb}|^2=1$, which corresponds to $|u_{bb}|=|u_{cb}|=1/\sqrt{2}$. Thus, the arbitrary choice of a balanced mode-swapping operation for our scheme is ultimately justified at least in the range of validity of the present analytical treatment. In the balanced configuration we obtain
\be
   \Delta^2(\theta_A-\theta_B)\big\vert_{\theta_A=\theta_B=\pi/2}= \frac{4e^{-2r}}{N}+\frac{4\bar{n}_s}{N^2}, \quad \bar{n}_s\ll N.
\ee
This is a function of $\bar{n}_s$ with a minimum at the point $\bar{n}_s=\sqrt{N}/2$, given by 
\be
    \min_{\bar{n}_s}\Delta^2(\theta_A-\theta_B)\big\vert_{\theta_A=\theta_B=\pi/2}=\frac{4}{N^{3/2}}, \quad \bar{n}_s\ll N.
\ee
Thus, the gain is found to have a peak corresponding to the value $\mathcal{G}_{\rm max}^2=\sqrt{N}$.

\subsection{Optimization of the $\hat{J}_z$ rotation for given values of the mode-swapping parameters}
\label{AppB2}

In this section, we consider Eq. \makeref{specific_sensitivity} and assume $\phi_{a,0}=\phi_{d,0}=0$, $\varphi_{0}=\pi/2$, $\delta_{bb}=0$ and  $|u_{bb}|=|u_{cb}|=1/\sqrt{2}$ right from the start, this time, getting:

\begin{align}
\label{specific_specific_sensitivity}
\Delta^2(\theta_A-\theta_B)\big\vert_{\theta_A=\theta_B=\pi/2}&= \frac{N}{\left(N-\bar{n}_s\right)^2}\Bigg\{\left(e^{2r}-1\right)\left[\sin{\left(\frac{\pi}{4}+\nu\right)}+\sin{\left(\frac{\pi}{4}+\frac{\nu}{2}+\delta\right)}\right]^2\nonumber\\
&-\left(1-e^{-2r}\right)\left[\cos{\left(\frac{\pi}{4}+\nu\right)}+\cos{\left(\frac{\pi}{4}+\frac{\nu}{2}+\delta\right)}\right]^2\Bigg\}\\
&+4\frac{N+\bar{n}_s}{\left(N-\bar{n}_s\right)^2}\nonumber,
\end{align}

where we have replaced $\chi_A$ and $\chi_B$ with their explicit expressions and set $\nu_E\equiv\nu$, $\delta_{cb}\equiv\delta$ for the sake of a lighter notation. We aim to study the \textit{absolute} minimum of the function
\beq\label{f_minimize}
f(\nu,\delta,r)&=&\left(e^{2r}-1\right)\left[\sin{\left(\frac{\pi}{4}+\nu\right)}+\sin{\left(\frac{\pi}{4}+\frac{\nu}{2}+\delta\right)}\right]^2\nonumber\\
&-&\left(1-e^{-2r}\right)\left[\cos{\left(\frac{\pi}{4}+\nu\right)}+\cos{\left(\frac{\pi}{4}+\frac{\nu}{2}+\delta\right)}\right]^2
\eeq
with respect to the variable $\nu$, for any fixed value of $\delta$ and of $r$. More specifically, we aim to determine the function $\nu_{\rm min}(\delta,r)$ giving the minimum point $\nu_{\rm min}$ for any desired value of $\delta$ and $r$. Notice that, due to the $4\pi$ periodicity of Eq. \makeref{f_minimize} as a function of $\nu$, infinitely many values can actually be attributed to $\nu_{\rm min}(\delta,r)$ for any fixed $\delta$ and $r$. Such degeneracy can be lifted, nonetheless, by arbitrarily restricting the range of $\nu_{\rm min}(\delta,r)$ to any given interval of length $4\pi$. We will do it below. Let us first highlight two important symmetry properties of $f(\nu,\delta,r)$.
\begin{itemize}
    \item $f(\nu+\pi+4n\pi,\delta+\pi/2,r)=f(\nu,\delta,r)$. As a consequence, the absolute minimum of $f(\nu,\delta,r)$ must be the same at $\delta$ as at $\delta+\pi/2$ and the minimum-point function possesses the property 
    \begin{equation}\label{first_symmetry}
        \qquad \qquad \nu_{\rm min}\left(\delta+\frac{\pi}{2},r\right)+4n\pi=\nu_{\rm min}(\delta,r)+\pi, \qquad n = 0, \pm 1, \dots.
    \end{equation}
    Thus, it is enough to determine this function for $\delta$ restricted to any interval of length $\pi/2$. Without loss of generality, we choose $\delta \in [-5\pi/8,-\pi/8[$, an interval of length $\pi/2$ centered at $\delta=-3\pi/8$. Concerning this and the following equation, notice that each different value of $n=0,\pm 1,\dots$ corresponds to a different choice for the range of $\nu_{\rm min}(\delta,r)$ so that a specific value can be assigned to $n$ once the degeneracy of $\nu_{\rm min}(\delta,r)$ has been lifted.
    \item $f(-3\pi/4-\Delta\nu+4n\pi,-3\pi/8-\Delta\delta,r)=f(-3\pi/4+\Delta\nu,-3\pi/8+\Delta\delta,r)$. As a consequence, the absolute minimum of $f(\nu,\delta,r)$ must be the same at $\delta=-3\pi/8-\Delta\delta$ as at $\delta=-3\pi/8+\Delta\delta$ and the minimum-point function possesses the property 
    \begin{equation}\label{second_symmetry}
        \qquad \qquad -\frac{3}{4}\pi+2n\pi-\nu_{\rm min}\left(-\frac{3}{8}\pi-\Delta\delta,r\right)=\nu_{\rm min}\left(-\frac{3}{8}\pi+\Delta\delta,r\right)+\frac{3}{4}\pi-2n\pi,  \qquad n = 0, \pm 1, \dots,
    \end{equation} 
    or, the graph of $\nu_{\rm min}(\delta,r)$ is -- for fixed $r$ and $n$ -- symmetric with respect to the point $(\delta,\nu_n)=(-3\pi/8,-3\pi/4+2n\pi)$. As a direct corollary of this property, the following prediction holds: should $\nu_n=-3\pi/4+2n\pi$ not be the minimum points corresponding to $\delta=-3\pi/8$, then the minimum-point function $\nu_{\rm min}(\delta,r)$ must be double-valued at $\delta=-3\pi/8$ and its two values symmetric with respect to $\nu_n$. The first possibility can be easily ruled out by noticing that $\nu_n=-3\pi/4+2n\pi$ are (absolute) maximum points of $f(\nu,\delta,r)$ for $\delta=-3\pi/8$ and a generic $r$. Indeed, the points $(\delta,\nu_n)=(-3\pi/8,-3\pi/4+2n\pi)$ simultaneously maximize the first squared term in Eq. \makeref{f_minimize} and minimize the second one. Hence, we must conclude that $\nu_{\rm min}(\delta,r)$ is double-valued at $\delta=-3\pi/8$, its two values being symmetric with respect to $\nu_n=-3\pi/4+2n\pi$.
\end{itemize}
On account of Eq. \makeref{ab_minimum_alpha_delta_2}, we also claim that
\begin{align}
    &\nu_{\rm min}\left(\delta=-\frac{5}{8}\pi,r\right)=\frac{11}{4}\pi\label{abs_minimum_point_1}\\
    &\nu_{\rm min}\left(\delta=-\frac{\pi}{8},r\right)=\frac{15}{4}\pi,\label{abs_minimum_point_2}
\end{align}
where the first and second equations were obtained for $l=-1,m=1$ and $l=0,m=1$, respectively. These specific values of the parameters were so chosen as to have $\nu_{\rm min}(\delta,r) \in [0,4\pi[$. Such condition lifts the degeneracy of the minimum-point function due to the periodicity of $f(\nu,\delta,r)$. It establishes $\nu_{\rm min}(\delta,r)$ as a well defined function of $\delta$ and $r$ and will always be  implicitly assumed in what follows. 

The two equations above are crucial in view of the complete determination of $\nu_{\rm min}(\delta,r)$. The missing link is the evaluation of the derivative of Eq. \makeref{f_minimize} with respect to $\nu$:
\be
    \frac{\partial}{\partial\nu}f(\nu,\delta,r)=4\cos{\left(\frac{\nu}{4}-\frac{\delta}{2}\right)}g(\nu,\delta.r),
\ee
where
\begin{align}
    g(\nu,\delta.r)=&\left(e^{2r}-1\right)\sin{\left(\frac{\pi}{4}+\frac{3}{4}\nu+\frac{\delta}{2}\right)}\left[\cos{\left(\frac{\pi}{4}+\nu\right)}+\frac{1}{2}\cos{\left(\frac{\pi}{4}+\frac{\nu}{2}+\delta\right)}\right]\nonumber\\
    +&\left(1-e^{-2r}\right)\cos{\left(\frac{\pi}{4}+\frac{3}{4}\nu+\frac{\delta}{2}\right)}\left[\sin{\left(\frac{\pi}{4}+\nu\right)}+\frac{1}{2}\sin{\left(\frac{\pi}{4}+\frac{\nu}{2}+\delta\right)}\right].
\end{align}
Vanishing of the partial derivative at the minimum points $(\delta,\nu)=(-5\pi/8,11\pi/4)$ and $(\delta,\nu)=(-\pi/8,15\pi/4)$, Eqs. \makeref{abs_minimum_point_1} and \makeref{abs_minimum_point_2} respectively, is due to the vanishing sine terms of $g(\nu,\delta,r)$. Indeed, with regard to the arguments of those sines, one has
\vspace*{-\baselineskip}
\begin{center}
\begin{tabular}{p{4cm}p{4cm}p{4cm}}
\begin{equation}\label{sine_factors_1_1}
    \frac{\pi}{4}+\nu=3\pi,
\end{equation}
&
\begin{equation}\label{sine_factors_1_2}
    \frac{\pi}{4}+\frac{\nu}{2}+\delta=\pi,
\end{equation}
&
\begin{equation}\label{sine_factors_1_3}
    \frac{\pi}{4}+\frac{3}{4}\nu+\frac{\delta}{2}=2\pi.
\end{equation}
\end{tabular}
\end{center}
\vspace*{-\baselineskip}
corresponding to Eq. \makeref{abs_minimum_point_1}, and
\vspace*{-\baselineskip}
\begin{center}
\begin{tabular}{p{4cm}p{4cm}p{4cm}}
\begin{equation}\label{sine_factors_2_1}
    \frac{\pi}{4}+\nu=4\pi,
\end{equation}
&
\begin{equation}\label{sine_factors_2_2}
    \frac{\pi}{4}+\frac{\nu}{2}+\delta=2\pi,
\end{equation}
&
\begin{equation}\label{sine_factors_2_3}
    \frac{\pi}{4}+\frac{3}{4}\nu+\frac{\delta}{2}=3\pi.
\end{equation}
\end{tabular}
\end{center}
\vspace*{-\baselineskip}
corresponding to Eq. \makeref{abs_minimum_point_2}. We then ask what happens to $\nu_{\rm min}(\delta,r)$ as we move away from the special values $\delta=-5\pi/8$ and $\delta=-\pi/8$. It seems reasonable to assume that, if the perturbation on $\delta$ is sufficiently small, the new minimum point $\nu_{\rm min}$ will be close to the old ones and Eqs. \makeref{sine_factors_1_1}-\makeref{sine_factors_2_3} still approximately correct. The latter is only true because we are focusing on the absolute minimima of $f(\nu,\delta,r)$: different stationary points of the function, including local minima, are not expected to satisfy the same equations (not even approximately), though still corresponding to a vanishing partial derivative.

Bringing all these observations together, let us first take $\delta \in [-5\pi/8-\epsilon,-5\pi/8+\epsilon]$ and assume $\epsilon$ is sufficiently small to apply the approximations $\sin{x} \approx 3\pi-x$, $\cos{x} \approx -1$, Eq. \makeref{sine_factors_1_1}, $\sin{x} \approx \pi-x$, $\cos{x} \approx -1$, Eq. \makeref{sine_factors_1_2}, and $\sin{x} \approx x-2\pi$, $\cos{x} \approx 1$, Eq. \makeref{sine_factors_1_3}, to the corresponding sine and cosine terms appearing in $g(\nu,\delta,r)$. Solving $g(\nu,\delta,r)=0$ for $\nu$ will then yield
\begin{align}\label{min_point_function_1}
    \qquad \qquad \nu_{\rm min}(\delta,r)&=\frac{\frac{21}{8}\pi\left(e^{2r}-1\right)+\frac{25}{8}\pi\left(1-e^{-2r}\right)-\left[\frac{3}{4}\left(e^{2r}-1\right)+\frac{1}{2}\left(1-e^{-2r}\right)\right]\delta}{\frac{9}{8}\left(e^{2r}-1\right)+\frac{5}{4}\left(1-e^{-2r}\right)},\\
    &\mathrm{with}\,\,\,\, \delta \in \left[-\frac{5}{8}\pi-\epsilon,-\frac{5}{8}\pi+\epsilon\right]\nonumber.
\end{align}
As a consistency check, one can verify that this correctly gives back Eq. \makeref{abs_minimum_point_1} if evaluated at $\delta=-5\pi/8$. In a similar way, taking $\delta \in [-\pi/8-\epsilon,-\pi/8+\epsilon]$ with $\epsilon$ sufficiently small to apply analogous approximations to the ones used above, but coming now from Eqs. \makeref{sine_factors_2_1}-\makeref{sine_factors_2_3}, and solving $g(\nu,\delta,r)=0$ for $\nu$ will give
\begin{align}\label{min_point_function_2}
    \qquad \qquad \nu_{\rm min}(\delta,r)&=\frac{\frac{33}{8}\pi\left(e^{2r}-1\right)+\frac{37}{8}\pi\left(1-e^{-2r}\right)-\left[\frac{3}{4}\left(e^{2r}-1\right)+\frac{1}{2}\left(1-e^{-2r}\right)\right]\delta}{\frac{9}{8}\left(e^{2r}-1\right)+\frac{5}{4}\left(1-e^{-2r}\right)},\\
    &\mathrm{with}\,\,\,\, \delta \in \left[-\frac{\pi}{8}-\epsilon,-\frac{\pi}{8}+\epsilon\right].
\end{align}
As in the previous case, Eq. \makeref{abs_minimum_point_2} is recovered for 
$\delta=-\pi/8$. 

\begin{figure}[t!]\includegraphics[width=14cm]{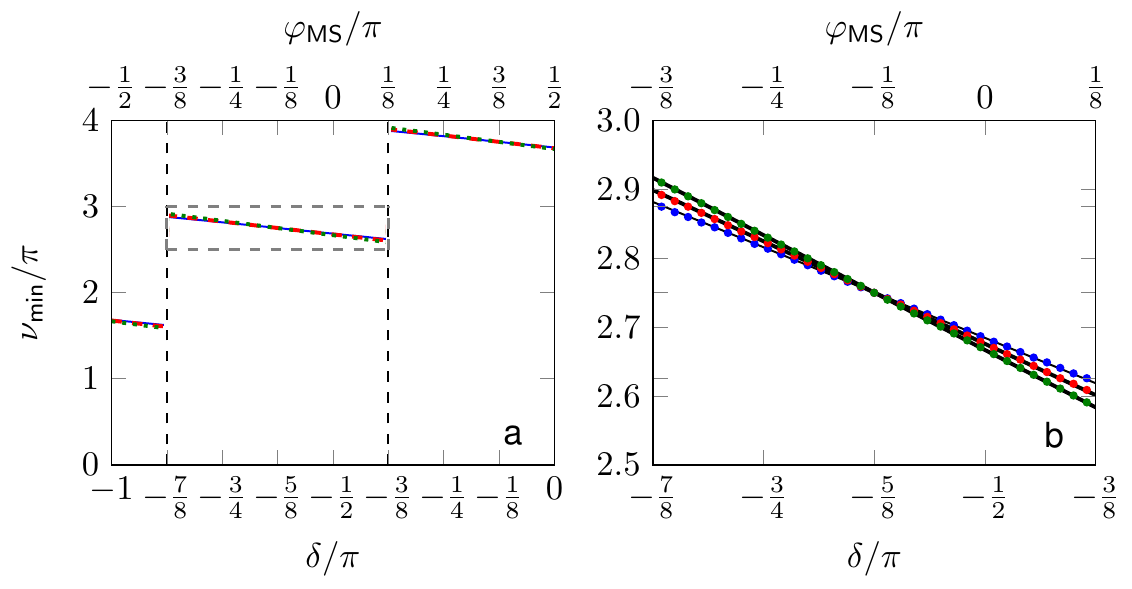}\caption{Plots of the minimum-point function $\nu_{\rm min}(\delta,r)$ giving the optimal rotation angle $\nu\equiv\m$ as function of $\delta\equiv\delta_{cb}$ and $r$. In panel (a), $\nu_{\rm min}(\delta,r)$ was evaluated by numerically minimizing Eq. \makeref{f_minimize} with respect to $\nu$ on the interval $\delta\in[-\pi,0]$ and for three representative values of $r$. The continuous blue line is relative to the small-$r$ regime, the dashed red line to the intermidiate regime and the dotted green line to the large-$r$ regime -- the three lines are better resolved in panel (b). The interval $\delta\in[-\pi,0]$ was specifically chosen to match the range $\varphi_{\rm MS}\in[-\pi/2,\pi/2]$ used for $\varphi_{\rm MS}$ in Fig. \ref{fig_4} (recall that $\delta=\varphi_{\rm MS}-\pi/2$). When comparing panel (b) of that figure with panel (a) of the present one, a good deal of agreement is found between the results of the analytical and the numerical approach. In particular, both figures show ``jumps" in the minimum-point function that clearly prove the validity of Eqs. \makeref{first_symmetry} and \makeref{second_symmetry}. In panel (b), we focus on the interval $\delta\in[-7/8\pi,-3/8\pi]$ and show that the results of the numerical minimization of Eq. \makeref{f_minimize} for the small, intermediate and large-$r$ regime are perfectly reproduced by Eqs. \makeref{3}, \makeref{1} and \makeref{5}, respectively (all plotted as continuous black lines). Thus, we can provide an analytical justification to the linear relation observed in Fig. \ref{fig_4} between the MS parameter $\varphi_{\rm MS}$ and the corresponding optimal rotation angle $\nu_E$.}
\label{figSupp_2}
\end{figure}

Last point to be addressed here is whether Eqs. \makeref{min_point_function_1} and \makeref{min_point_function_2} provide or not a good approximation to $\nu_{\rm min}(\delta,r)$ as $\delta$ is strongly perturbed away from the special points
$\delta=-5\pi/8$ and $\delta=-\pi/8$. In Fig. \ref{figSupp_2}, panel (b), we show a direct comparison between Eq. \makeref{min_point_function_1} (black continuous lines) and the result of a numerical evaluation of $\nu_{\rm min}(\delta,r)$ (colored dots).
The approximation was tested on an interval of length $\pi/2$ centered at $\delta=-5\pi/8$, for three values of the parameter $r$, and found to always fit the numerical data with high precision. The same can be said about Eq. \makeref{min_point_function_2} if tested on an interval of length $\pi/2$ centered at $\delta=-\pi/8$: more in general and according to the symmetry property highlighted in Eq. \makeref{first_symmetry}, the minimum-point function is fully determined, as a function of $\delta$, by its behavior on any interval of length $\pi/2$. As a final remark, notice that the three values of $r$ used for the test were so chosen as to sample all the possible regimes relative to that parameter. Indeed, three distinct regimes can be identified: $(i)$ small-$r$ regime (short squeezing time), in which $e^{2r}-1 \ll 1$, $1-e^{-2r} \ll 1$ and we are allowed to take $e^{2r} -1 \approx 2r \approx 1 - e^{-2r}$; $(ii)$ large-$r$ regime (long squeezing time), in which $e^{2r} \gg 1 \gg e^{-2r}$; $(iii)$ an intermediate regime, for which none of the previous assumptions is suitable. Both $(i)$ and $(ii)$ correspond to special limiting forms of our equations, which become $r$-independent both in the small and large-$r$ regimes. In the following list of equations we sum up the relevant information found about $\nu_{\rm min}(\delta,r)$. 
\begin{align}
    \nu_{\rm min}(\delta,r)&=\frac{\frac{21}{8}\pi\left(e^{2r}-1\right)+\frac{25}{8}\pi\left(1-e^{-2r}\right)-\left[\frac{3}{4}\left(e^{2r}-1\right)+\frac{1}{2}\left(1-e^{-2r}\right)\right]\delta}{\frac{9}{8}\left(e^{2r}-1\right)+\frac{5}{4}\left(1-e^{-2r}\right)}, \,\,\, -\frac{7}{8}\pi \leqslant \delta \leqslant -\frac{3}{8}\pi,\label{1}\\
    \nu_{\rm min}(\delta,r)&=\frac{\frac{33}{8}\pi\left(e^{2r}-1\right)+\frac{37}{8}\pi\left(1-e^{-2r}\right)-\left[\frac{3}{4}\left(e^{2r}-1\right)+\frac{1}{2}\left(1-e^{-2r}\right)\right]\delta}{\frac{9}{8}\left(e^{2r}-1\right)+\frac{5}{4}\left(1-e^{-2r}\right)}, \,\,\, -\frac{3}{8}\pi\leqslant \delta \leqslant \frac{\pi}{8}.\label{2}
\end{align}
\begin{itemize}
    \item small-$r$ limit:
    \begin{align}
    \nu_{\rm min}(\delta,r)&= \frac{46}{19}\pi-\frac{10}{19}\delta \qquad -\frac{7}{8}\pi \leqslant \delta \leqslant -\frac{3}{8}\pi,\label{3}\\
    \nu_{\rm min}(\delta,r)&=\frac{70}{19}\pi-\frac{10}{19}\delta \qquad -\frac{3}{8}\pi\leqslant \delta \leqslant \frac{\pi}{8}.\label{4}
    \end{align}
    \item large-$r$ limit:
    \begin{align}
    \nu_{\rm min}(\delta,r)&= \frac{7}{3}\pi-\frac{2}{3}\delta \qquad -\frac{7}{8}\pi \leqslant \delta \leqslant -\frac{3}{8}\pi,\label{5}\\
    \nu_{\rm min}(\delta,r)&=\frac{11}{3}\pi-\frac{2}{3}\delta \qquad -\frac{3}{8}\pi\leqslant \delta \leqslant \frac{\pi}{8}.\label{6}
    \end{align}
\end{itemize}

\section{Comparison with numerical results}
\label{AppC}

\begin{figure}[h!]
\includegraphics[width=12cm]{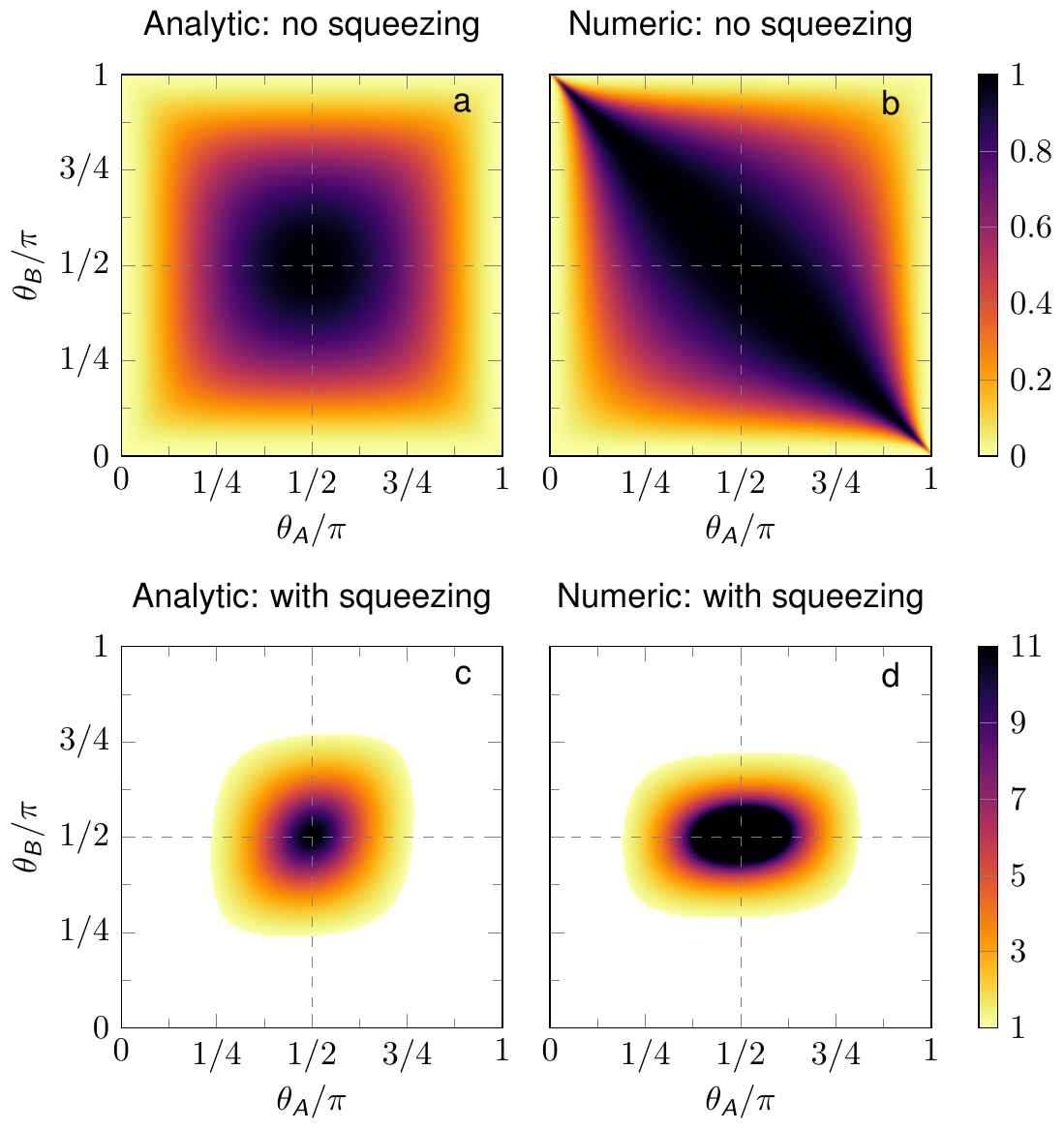}
\centering
\caption{Evolution of the sensitivity gain, $\mathcal{G}^2(\theta_A,\theta_B)$, in color plot, with a comparison between the result of the analytical model (a-c) and the numerical method (b-d). Panels (a) and (b) show the case of a vanishing squeezing time $\tau_E=0$, while panels (c) and (d) shows $\tau_E \neq 0$ with a squeezing time corresponding to the maximum gain.}
\centering
\label{figSupp_3}
\end{figure}

Due to the approximations A1 and A2 introduced in Eqs. \makeref{step1} and \makeref{psi_s}, respectively, the estimation uncertainty obtained from our analytical treatment cannot perfectly match the numerical results. At the optimal working point, $\theta_A=\theta_B=\pi/2$, A1 has no effect, as proved by the fact that the analytical and the numerical approach reach identical shot-noise sensitivity for a vanishing amount of squeezing. To the contrary, small to quite large discrepancies between the analytical and the numerical method can be found for generic values of the phase shifts $\theta_A, \theta_B$ even when a vanishing amount of squeezing is considered. Being unrelated to squeezing, these have to be attributed to the first approximation A1. Let us develop this point further. To start with, we consider Eq. \makeref{specific_specific_sensitivity} completed with the addition of its phase-dependent part $Q(\cot{\theta_A},\cot{\theta_B})$:
\begin{align}
\Delta^2(\theta_A-\theta_B)=& \frac{N}{\left(N-\bar{n}_s\right)^2}\left(e^{2r}-1\right)\left[\sin{\left(\frac{\pi}{4}+\nu\right)}+\sin{\left(\frac{\pi}{4}+\frac{\nu}{2}+\delta\right)}\right]^2\nonumber\\
-&\frac{N}{\left(N-\bar{n}_s\right)^2}\left(1-e^{-2r}\right)\left[\cos{\left(\frac{\pi}{4}+\nu\right)}+\cos{\left(\frac{\pi}{4}+\frac{\nu}{2}+\delta\right)}\right]^2\\
+&4\frac{N+\bar{n}_s}{\left(N-\bar{n}_s\right)^2}+2\frac{N+\bar{n}_s}{\left(N-\bar{n}_s\right)^2}\left(\cot^2{\theta_A}+\cot^2{\theta_B}\right)+\frac{\left(2\bar{n}_s+1\right)\bar{n}_s}{\left(N-\bar{n_s}\right)^2}\left(\cot{\theta_A}-\cot{\theta_B}\right)^2\nonumber,\label{specific_specific_sensitivity_with_Q}
\end{align}
The uncertainty obtained for a vanishing squeezing time $\tau_E=0$ (corresponding to both $r=0$ and $\bar{n}_s=0$) is 
\begin{equation}\label{no_squeezing_1}
    \Delta^2(\theta_A-\theta_B)\vert_{\rm th}=\frac{2}{N}\left[\cot^2{\theta_A}+\cot^2{\theta_B}+2\right].
\end{equation}
By comparison, the outcome of the numerical method, which can be \textit{exactly} predicted for the case of no squeezing, is found to be:
\begin{align}
    \Delta^2(\theta_A-\theta_B)\vert_{\rm num}=\Delta^2(\theta_A-\theta_B)\vert_{\rm th}-\frac{1}{N}\left(\cot{\theta_A}-\cot{\theta_B}\right)^2\label{no_squeezing_2}.
\end{align}
With no squeezing applied, the analytical approach generally overestimates the uncertainty of our scheme, as $\Delta^2(\theta_A-\theta_B)\vert_{\rm num}\leqslant\Delta^2(\theta_A-\theta_B)\vert_{\rm th}$. This discrepancy is caused by the second term in Eq. \makeref{no_squeezing_2}, namely $\left(\cot{\theta_A}-\cot{\theta_B}\right)^2/N$. For $\theta_A=\theta_B$ one has $(\cot{\theta_A}-\cot{\theta_B})^2=0$, the extra term is minimum (vanishing) and $\Delta^2(\theta_A-\theta_B)\vert_{\rm num}=\Delta^2(\theta_A-\theta_B)\vert_{\rm th}$. For $\theta_B=\pi-\theta_A$, the difference $\Delta^2(\theta_A-\theta_B)\vert_{\rm th}-\left(\cot{\theta_A}-\cot{\theta_B}\right)^2/N$ turns out to be independent of $\theta_A$ and $\theta_B$, giving a constant estimation uncertainty $\Delta^2(\theta_A-\theta_B)\vert_{\rm num}=4/N$ that attains its minimum, at the shot-noise limit. All these details are visualized in Fig. \ref{figSupp_3}.

The gain functions $\mathcal{G}^2(\theta_A,\theta_B)=4\Delta^2(\theta_A-\theta_B)/N$ corresponding to Eqs. \makeref{no_squeezing_1} and \makeref{no_squeezing_2} have been plotted in panel (a) and (b), respectively. Looking at the figure, it becomes clear that the main effect of the term $\left(\cot{\theta_A}-\cot{\theta_B}\right)^2/N$ is to break the left-right and up-down symmetries of Eq. \makeref{no_squeezing_1}, panel (a), and to enhance the sensitivity gain along the direction $\theta_B=\pi-\theta_A$. Finally, it is worth mentioning what the effect of a non-vanishing squeezing is for each method. As can be seen from Eq. \makeref{specific_specific_sensitivity_with_Q}, a non-vanishing squeezing time $\tau_E\neq 0$ (implying $\bar{n}_s\neq 0$ in the sensitivity equation) is again related to the emergence of a term $(\cot{\theta_A}-\cot{\theta_B})^2$ which is, this time, not subtracted from but added to the sensitivity. As a consequence, the gain is now mainly enhanced along the $\theta_B=\theta_A$ direction. In this case, the general behavior of the gain function resulting from the numerical method is similar to that obtained from the analytical prediction [compare panel (c) and (d) in Fig. \ref{figSupp_3}]. 

\section{Two-Axis Twisting and Quantum Non-Demolition measurements}
\label{AppD}

\begin{figure}[h!]\includegraphics[width=\textwidth]{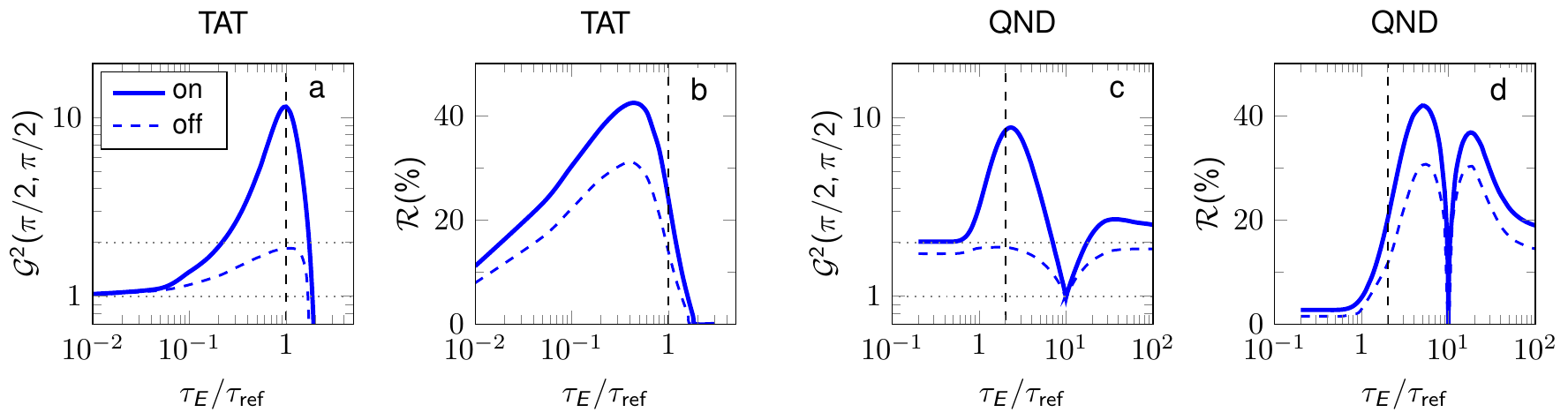}
\caption{Maximum sensitivity gain, $\mathcal{G}^2(\pi/2,\pi/2)$, and bandwidth, $\mathcal{R}$, as a function of the squeezing strength $\tau_E$ for different entanglement strategies. In each panels, the thin blue solid line are numerical results obtained for total $N=100$ particles. The blue dashed lines refer to the separable differential scheme using a coherent spin state ($\tau_S^B=0$) a spin squeezed state ($\tau_S^A=\tau_E$) and no MS. (a\&b) Two-Axis Twisting (TAT) with $\tau_{ref}=log_{10}(2\pi N)/(4N)$. (c\&d) Quantum Non-Demolition measurements (QND) with $\tau_{ref}=1$.}
\label{figSupp_1}
\end{figure}

In Fig~\ref{figSupp_1}, we plot the sensitivity gain, $\mathcal{G}^2$, at the optimal working point and the bandwidth, $\mathcal{R}$, as a function of the squeezing time $\tau_E/\tau_{\rm ref}$ in the case where the squeezing dynamic is generated through Two-Axis Twisting (TAT) dynamic [Fig~\ref{figSupp_1}a,b] and Quantum Non-Demolition (QND) measurements [Fig~\ref{figSupp_1}c,d]. 
Following Ref.~\cite{KitagawaPRA1993}, the state in Eq.~\ref{Eq_State_step_2_OAT} is now replace by 
\be
\ket{\psi_2} = \exp\left\{-i\m \Jz{a,b}\right\} \otimes \exp\left\{- \tau_E \left[\left(\hat{J}_+^{\{a,b\}}\right)^2-(\hat{J}_-^{\{a,b\}})^2\right]\right\} \ket{\psi_1},
\ee
in the case of TAT where $\hat{J}_{\pm}^{\{a,b\}}=\hat{J}_x^{\{a,b\}}\pm i\hat{J}_y^{\{a,b\}}$ and by a Gaussian state~\cite{PezzePRX2021} in the case where entanglement is generated by QND protocol.
The results are shown in the case where the number of particles is fixed to $N=100$.
The two protocols reproduce a similar behavior than the one shown in Fig.\ref{fig_2} with One-Axis-Twisting dynamic. Because the orientation of the state is always optimized, QND measurements is characterized by two maximum. 
Here we conclude that the mode-swapping method can be extended to various squeezing procedures and is not limited to One-Axis-Twisting dynamic.

\begin{thebibliography}{100}
\bibitem{notaSQL}
This bound is obtained considering the relation $\Delta^2 (\theta_A - \theta_B) = \Delta^2 \theta_A + \Delta^2 \theta_B$, valid for independent interferometers, and taking coherent spin states of $N_A$ and $N_B$ particles, respectively, such that 
$\Delta^2 \theta_{A,B}=1/N_{A,B}$, independently from the value of $\theta_{A,B}$.
Finally, the optimal separable configuration is obtained for $N_A=N_B=\NT/2$, giving $\Delta^2 (\theta_A - \theta_B)_{\rm SQL}=4/N$.
\bibitem{notaSS}
We take a entangled state of $N_A$ particles and an coherent spin state of $N_B = \NT - N_A$ particles in the interferometers $A$ and $B$, respectively. 
For the mode-separable case we have $\Delta^2 (\theta_A - \theta_B) = \Delta^2 \theta_A + \Delta^2 \theta_B$.
Let's suppose that $\Delta^2 \theta_A \ll \Delta^2 \theta_B=1/N_B$. 
The optimization of $\Delta^2 (\theta_A - \theta_B)$ with respect to $N_A$, gives $\Delta^2 (\theta_A - \theta_B) \sim 1/\NT$.
Instead, if two interferometers have the same number of particles, $N_A = N_B = \NT/2$, we obtain $\Delta^2 (\theta_A - \theta_B) \sim 2/\NT$.
\end{thebibliography}
\end{document}